\documentclass[a4paper,11pt]{article}
\pdfoutput=1 

\usepackage{jheppub} 

\usepackage[T1]{fontenc} 
\usepackage{url}
\usepackage{comment}
\usepackage{color,soul}
\usepackage{subfig}
\usepackage{tabularx}
\usepackage{graphicx}
\usepackage{amsmath}
\usepackage{amsfonts}
\usepackage{amssymb}
\usepackage{comment}

\title{\boldmath Jet characterization in Heavy Ion Collisions by QCD-Aware Graph Neural Networks}

\author[a,1]{Yogesh Verma,\note{Corresponding author.}}
\author[a,2]{Satyajit Jena,\note{Corresponding author.}}

\affiliation[a]{Department of Physical Sciences, Indian Institute of Science Education and Research (IISER) Mohali, Sector 81 SAS Nagar, Manauli PO 140306 Punjab, India}

\emailAdd{ms16027@iisermohali.ac.in}
\emailAdd{sjena@iisermohali.ac.in}

\abstract{

The identification of jets and their constituents is one of the critical problems and challenging tasks in heavy-ion experiments such as experiments at RHIC and LHC. The huge background of soft particles poses a curse for jet-finding techniques. The inabilities or lack of efficient techniques to filter out the background lead to a fake or combinatorial jet formation which may have an erroneous interpretation. This article presents the GraphReduction technique (GraphRed), a novel class of physics-aware attention graph neural networks built upon jet physics in heavy-ion collisions. Since, the number of tracks are expected to vary from one event to another, this approach finds the most likely jet constituent particles on an event-by-event basis. This technique demonstrates the robustness and applicability of this method for finding jet constituent particles and shows the applicability of graph architectures on particle-level classification in each heavy-ion event produced at the LHC.

}

\begin{document} 
\maketitle
\section{Introduction}
\label{sec:intro}

The Quark Gluon Plasma (QGP) is created and studied by colliding ultra-relativistic heavy-ions at collider experiments such as Relativistic Heavy-Ion Collider (RHIC), Large Hadron Collider (LHC), etc. In the QGP state, the constituent quarks and gluons, partons, get deconfined due to the extreme energy densities. The energy density of such a system is expected to reach $0.2-1$ $GeV/fm^{3}$~\cite{Bazavov_2014,Karsch2002}. The partons which undergo the hard-scattering process early in the collision traverse through QGP medium, then fragments and hadronize into a collimated spray of particles called Jet. Due to color charges, these partons are expected to interact with the QGP medium and lose their energy through a combination of bremsstrahlung and collisional energy losses. The energy loss of these high momentum jet particles due to the collision with the medium is called jet quenching. Since, the jet quenching can happen only in the QCD medium; thus the study of the jet in the heavy-ion collision can also serve as one of the signatures of QGP. The jet quenching results in modification of the properties of the resulting jet, and the lost energy partially retains its correlation with the parent partons. However, high $p_{T}$ particles dominate the jet, but these low momentum particles carry critical information about the parent parton interaction with the medium. Although the jets are studied extensively in pp collisions~\cite{jet1,jet2,jet3}, the study of jets becomes extremely complicated due to the high multiplicity environment in heavy-ion collisions. An enormous amount of theoretical and experimental effort has been made around the treatment of jets and in medium modifications to develop techniques that map theoretical predictions with experimental data while constraining some inherent properties. The primary motivation for the jet study is to yield a calibrated probe of perturbative QCD (pQCD) by providing measurements of observables with a cross-section that can be compared to its calculations. It is worth observing that many physical observables are responsive to non-perturbative effects like hadronization, thus affecting the measurements of some observables such as momentum spectra, transverse energy, flow, etc. Therefore, a detailed study of jets to characterize the shape of the jet, constituent particle composition, fragmentation and splitting functions, etc. In the environment of large and complicated backgrounds in heavy-ion events depends on efficient jet-finding algorithms.

Most jet-finding algorithms work on sequential recombination of final state particles to form jets. The interactions of the hard partons and its daughter particles complicate the selection procedure of what should belong to a jet. This ambiguity makes the study of jets and interpretation of their result more challenging in heavy-ion collisions.  In addition, this gives rise to a finite contribution towards background to jets structure. The most significant source of background is due to the collective flow of the particles originating from the QGP medium. Methods~\cite{Connors_2018, 2011,Abelev_2012,2013220, Berta_2014} which are employed to suppress and subtract the background from events depends on the nature of assumptions about the background, which may change the sensitivity towards detection of in-medium modifications. These techniques may also bias the measured jet sample, for instance, by selecting gluon jets at a higher rate than quark jets.  For the jets in the heavy-ion collision, these analysis-cuts serve as the definition of the jet and can not be disregarded. It is often assumed about the hidden interplay between hard and soft induced particles and about the distribution of background. In experiments, the final state particles are measured and observed, which carries the traces of jets.  The jets are determined using a recombination algorithm~\cite{Salam_2010} from final state particles. These reconstructed jets are further used as the physics observable for characterizing the event shape and physics. These reconstruction techniques/algorithms are often mathematically challenging and computationally expensive tasks mainly due to the complex event structure. Such complex tasks can be tackled using advanced computational techniques like imaging, machine learning, etc. In that context, numerous machine learning approaches have been developed upon the calorimeter-level considering them as images ~\cite{Cogan_2015, de_Oliveira_2016, almeida2015playing, Baldi_2016,Guest_2016, Barnard_2017, Komiske_2017,Kasieczka_2017,verma2021shower}
and using recursive networks~\cite{Louppe_2019}. The graph networks have been proved to be efficient in various tasks ranging from tracking~\cite{verma2020particle} to jet finding~\cite{dreyer2020jet}. The general approach reduces the original jet classification to an image classification problem in the $\eta - \phi$ plane. Most of these works deal with developing algorithms for tagging and characterizing jets in mostly \textit{pp} collision events where backgrounds are comparatively smaller than \textit{AA} collision. At the same time, when we scale \textit{pp} collisions to \textit{AA} collisions, there is a subsequent change in the background distribution and signal (jets) distribution. In contrast to the \textit{pp} collisions there is a large background present in \textit{AA} collisions. These developed and stated algorithms had not been tested, applied, and constructed in keeping the \textit{AA} collisions large background in consideration.

This work proposes a solution to the large background problem prevalent in heavy-ion collisions based on the physics aware and attention graph neural networks. Foremost, we propose to embed the full final state particle content in an event into space, considering each particle as a point cloud with features associated with it by geometrical embedding, creating a web-shaped graph. This particle-level embedding can be fed into the graph neural network for particle-level classification into hydro (soft) and jet constituent particles (hard). This represents the first model operating on characterizing jets in heavy-ion events in lieu of large background distribution. We have studied our proposed method on monte-carlo simulated data and verified the proof of concept of our method in this manuscript. The manuscript is organized as follows: section~\ref{sec:problem} describes the classification and prediction tasks at the event level,  section~\ref{sec:Data} gives detail about model architecture, and section~\ref{sec:result} describes the results of our experiments on particle and jet-level and comparison with current frameworks. The paper concludes with Section~\ref{sec:conc}.

\section{Problem statement}
\label{sec:problem}

A typical heavy-ion event produces a large number of out-going particles which fly outwards from its interaction point. The events can be classified into several centrality classes depending on the initial collision geometry defined in terms of impact parameters. The small impact parameter is considered a central collision, whereas the large impact parameter is classified as peripheral collision. Such outgoing particles in a heavy-ion collision event can be described by varying number of final state particles  $N_{part} = [n_{i}|i=1,2...N]$ indexed by $i$, where each $n_{i}$ is represented by its 4 momentum vector $v_{i} \in \mathbb{R} $. 
The emitted particles are often represented by $y$ and $\phi$ coordinate pairs with respect to emission angle from the interaction point. $\phi$ is the azimuthal angle of emitted particle, and $y$ is the rapidity of a particle. Out of all outgoing particles in an event, isolating the hard particles becomes more cumbersome due to many particles. The task of jet-finding methods is to identify particles that are created through jet processes. Usually, the 4-momenta of particles are clustered using sequential recombination jet algorithm like $anti-k_{T}$~\cite{Cacciari_2008} by recursively combining the particles that minimize.

 \begin{equation}
\label{eq:edge}
d_{ij} = min(1/p_{T,i}^{2},1/p_{T,j}^{2})\frac{\Delta R_{ij}^{2}}{R^{2}} 
\end{equation}

where $\Delta R_{ij}^{2}$ $=$ $(y_{i}-y_{j})^{2} +(\phi_{i} - \phi_{j})^{2}$ and $R$ is the radius of jet. Contrary to the \textit{pp} collisions, \textit{AA} collision suffer from a large background of soft particles originating from QGP. Therefore, it must be acknowledged that the jets reconstructed from the jet-finding algorithm are not necessarily be generated only from hard processes. Thus, utmost care must be taken to interpret the results from the jet-finding algorithm to distinguish between fake or combinatorial jets to actual jets. It may also happen that particles correlated through a hard process are grouped into jet candidates with background particles, creating a difficult trade-off on identifying the actual jets and fake jets to improve the quality of jet characterizations in heavy-ion collisions.

This work builds on the above intersection by detecting the particles produced by hard processes in the event-level structure. Since a huge background is expected in the heavy-ion collision, our goal is to build an intermediate model between experimental event-level output particles and jet finding algorithm to remove the background footprint from the event. Before we discuss our method's detailed description, we would like to mention the dataset used for this analysis in the following section.

\section{Dataset generation}
\label{sec:gen}

There are several event generators available to simulate and study heavy-ion collision. In order to quantify the effectiveness of subtracting underlying background in a heavy-ion event, we need access to the background and hard jets via monte-carlo simulation. This can be either done by (i) simulating a hard $pp$ event in isolation and then embedded in a heavy-ion event background or (ii) simulation of a full heavy-ion event including jets by a monte-carlo event generator. In the first case, embedding a hard jet event over the background may lead to a loss in interactions of the hard partons and its daughter particles with strongly interacting QCD medium, which can result in a bias. We have applied our developed algorithm to both cases and compared the results with standard benchmark techniques to overcome this.

For generating the full event with jet production, We have used HYDJET++~\cite{hydjet} monte-carlo event generator for our study because (a) it provides detailed information about jet and its constituent particles and (b) it also provides the information about jet quenching. Thus, it becomes a natural choice to test and validate our model. The HYDJET++  uses PYTHIA~\cite{pythia} as a backend to generate parton shower and PYQUEN~\cite{pyquen} model that modifies the standard jet event into the complete event structure. The soft part is the thermal hadronic state generated on freeze-out hypersurface on chemical and thermal freeze-out obtained from relativistic hydrodynamics. For this study, we have generated \textit{PbPb} events at $\sqrt{s_{NN}}=2.76$ TeV at $0-5\%$ centrality (most central). Since jets originating from heavy-ion collisions can be quenched or non-quenched, we trained and evaluated our model on both quenched and non-quenched jets simulated datasets. For our embedding setup, we generated hard jet events using PYTHIA~\cite{pythia}; the background events are generated by HYDJET++~\cite{hydjet} at at $\sqrt{s_{NN}}=2.76$ TeV at $0-5\%$ centrality (most central). To summarize, we have listed the types of datasets generated for our study below:
\begin{itemize}
    \item Full heavy-ion events by HYDJET (incl. jets)
    \begin{itemize}
        \item Quenched dataset (quenching = ON)
        \item Non-Quenched dataset (quenching = OFF)
            \end{itemize}
    \item Embedded events by PYTHIA (jets) $\&$ HYDJET (background)

\end{itemize}
The above simulations provides us with various physical observables such as E (energy), $p_{x},p_{y},p_{z}$ (momentum components) and $x,y,z$ (coordinates). In general, an event generator doesn't  give a complete information of track of the particle and provides the starting point of the track of the particle. A feature vector of 7 variables i.e. $( E,p_{x},p_{y},p_{z},x,y,z)$ is constructed for each of the final state particles for geometrical embedding which is then served as a node features of the graph. Since both the simulation procedures and event generators provide us the explicit details of all the hard and soft particles, which is used to train our model.

\section{GraphRed model}
\label{sec:Data}
\begin{figure}[!hbt]
\def\tabularxcolumn#1{m{#1}}
\begin{tabularx}{\linewidth}{@{}cXX@{}}
\begin{tabular}{cc}
\subfloat[ XYZ system 
]{\hspace{-3.0em}\includegraphics[scale=0.33]{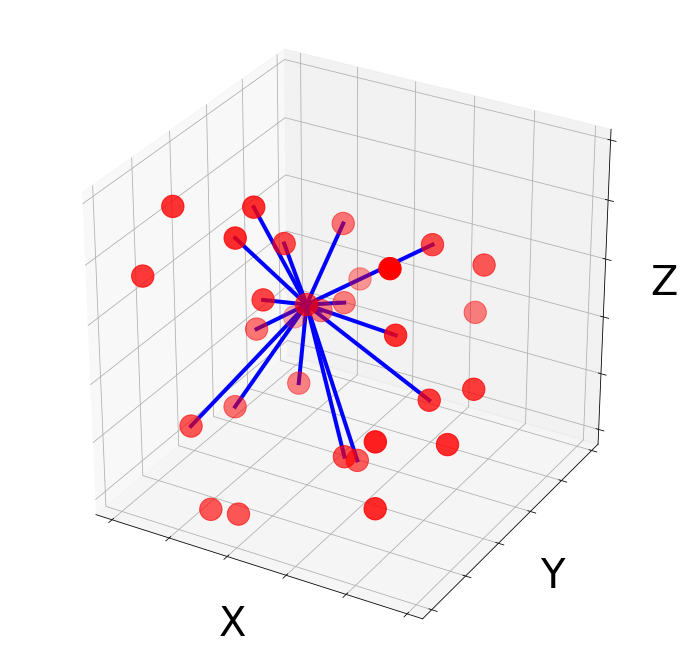}} 
   & \subfloat[($\eta$, $\phi$) plane]{\hspace{-1.0em}\includegraphics[scale=0.41]{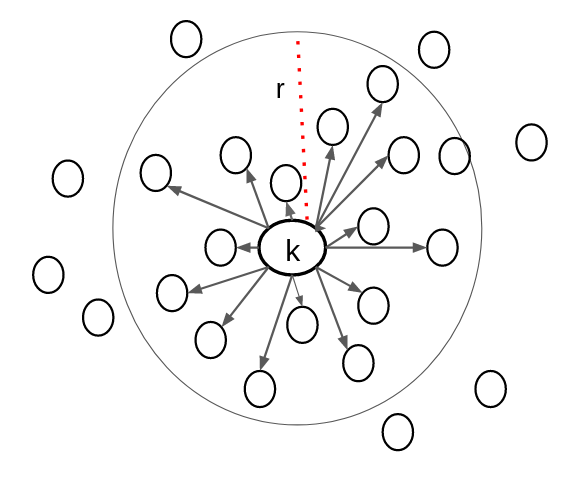}}\\
\end{tabular}
\end{tabularx}
\caption{(color online) Pictorial representation of web like graph construction for \textit{k}th particle in (X, Y, Z) system and ($\eta$, $\phi$) plane }\label{fig:geo}
\end{figure}

The GraphRed model is based on Attention-based Graph Neural Network (AGNN), which proposes a modification to previously applied Message Passing Neural Networks (MPNN) described in Ref.~\cite{kipf2017semisupervised,choma2020track}. The Attention-based mechanism has been successfully used in various tasks such as machine translation~\cite{bahdanau2016neural,vaswani2017attention}, machine reading~\cite{cheng2016long} and many more~\cite{graph}. An example of a single graph attention layer is given in ~\cite{graph} where an arbitrary graph attention network is constructed, and the layer computes the coefficient in the attention mechanism for the node pair. They have also proved to outperform Graph Convolutional Network (GCN)~\cite{kipf2017semisupervised} in several tasks, such as semi-supervised node classification, link prediction, and so on.\\
The entire process follows four distinct steps, (i) event generation for training and validation using HYDJET monte-carlo generator, (ii) starting an embedding procedure to create an input graph for the model, (iii) encoding the graph input in the model via various networks to learn appropriate features of the event generated in HYDJET and (iv) finally prediction of the type of the particle (hard or soft) and eventually reconstructing the jets. The event generation is discussed in detail in sect.~\ref{sec:gen}.  To create a graph serving as an input to the model, we devise a new strategy to embed particles in the multi-dimensional space named as Geometrical Embedding. The main objective of executing geometrical embedding is to represent the particle-level features in multi-dimensional space to exploit the latent representations. For each particle in an event, a web-like graph is constructed as shown in Fig.~\ref{fig:geo} where edges (arrows) are the connections between individual particles and the nodes (circles) are the individual particles.

The number of connections a particle can have is restricted by the \textit{r} parameter which is a measure of distance between each particle in ($\eta$, $\phi$) plane. The mathematical expression is 

 \begin{equation}
\label{eq:r}
r_{ij} = \sqrt{(\eta_{i} - \eta_{j})^{2} + (\phi_{i} - \phi_{j})^{2}}
\end{equation}

Each node consists of the particle-level feature vector of energy ($E$), momentum vector ($p_{x}, p_{y}, p_{z}$), position ($x, y, z$), which qualitatively transforms the embedding into a topological space of seven dimensions where each node has a definite projection in each plane and edges are connections in this multi-dimension space. These features and connections are the input to our GraphRed model that predicts the type of each particle by evaluating the connections between nearby particles in a multi-dimensional hyperspace. Once the geometrical embedding is executed and an input graph is created, it is passed through the encoder network to encode it into a latent representation.

The next step is to train the model with embedded input graphs which is done by following network architectures:
\begin{itemize}
    \item{\underline{Encoder Network}:} Multi layer Perceptron (MLP) consisting of single layer having 32 nodes whose input features are 7 and output features are 32 with layer normalization and  $\tanh$ activation function. 
    \item{\underline{Edge Network}:} It is a MLP consisting of 4 layers. The first three layers have 32 nodes where as last layer has only one node. Layer normalization and $\tanh$ activation function are used after first three layers.
    \item{\underline{Node Network}:} It is a MLP consisting of 4 layers. All layers have 32 nodes with layer normalization and  $\tanh$ activation function.
    \item{\underline{Node Classification Network}:} It is a MLP which consists of 2 layers having 32 and one node respectively. ReLU activation function is used after the first layer and sigmoid after the second layer.
\end{itemize}

 The input graph is passed through the encoder network to encode it into a latent representation. This Latent representation (output of encoder network) of data is then passed through a recurrent set of $\{$Node Network, Edge Network$\}$ where data passes from Node Network to Edge Network and then again to Node Network $n_{iter}$ times. Each time the data passes through Edge Network, edge features are computed for edges in the graph along with a score for each edge as the network's output. This score acts as an attention mechanism in the network. This score with the edge features is passed to Node Network, where node features are calculated. This whole cycle repeats by $n_{iter}$ times. After the cycle has been completed, the collective features of the graph are passed through the Node Classification network to compute the probability of each particle (node) belonging to the soft particle or the hard particle.

The GraphRed classification model described above performs binary classification of nodes of the graph to characterize the type of particle. The model uses four graph recursions of the network, which is followed by a final classification layer that evaluates whether the node belongs to the soft or hard type. Given the above architecture, we present the results of node classification to find the hard particles.  We present a hierarchical application of our graph network starting from particle level comparison to the jet level comparison and comparison with other background subtraction and reconstruction methods.

\section{Results and Discussion}
\label{sec:result}

\subsection{Training and Validation Results}
\subsubsection{Full heavy-ion event by HYDJET}
\label{sec:hydjet_train}
We started our procedure by training both quenched and non-quenched monte-carlo datasets with 100,000 events each. The model is trained over quenched and non-quenched datasets by collectively embedding each event's final state particles into its geometrical embedding. Then the geometrical embedded particles, event graphs, are passed through the networks for training. The training was performed for 50 epochs (epochs are the number of iterations of network processing).  The training and validation sets are divided in a ratio of $80:20$. The value of $r$ ($r$ is a distance parameter that defines the number of connections of each node in the network) was chosen wisely to be $0.6$. It is high enough to find all jet features and low enough to be computationally feasible. Usually, the high value of $r$ causes a large number of connections in the network leading to high computational demand.
On the other hand, a low value of $r$ causes fewer connections leading to inadequate feature evaluation. The model's performance is evaluated by an objective function called loss function, which determines the difference between the predicted and target outputs. The loss function is used to learn the model parameters such as weights and biases of neural networks by minimizing them with the help of an optimizer. We have used the binary loss function called binary cross-entropy loss and the ADAM optimizer~\cite{adam} as optimizer. 

\begin{figure}[!hbt]
\def\tabularxcolumn#1{m{#1}}
\begin{tabularx}{\linewidth}{@{}cXX@{}}
\begin{tabular}{cc}
\subfloat[ROC for Non-Quencehd dataset]{\hspace{-1.0em}\includegraphics[scale=0.18]{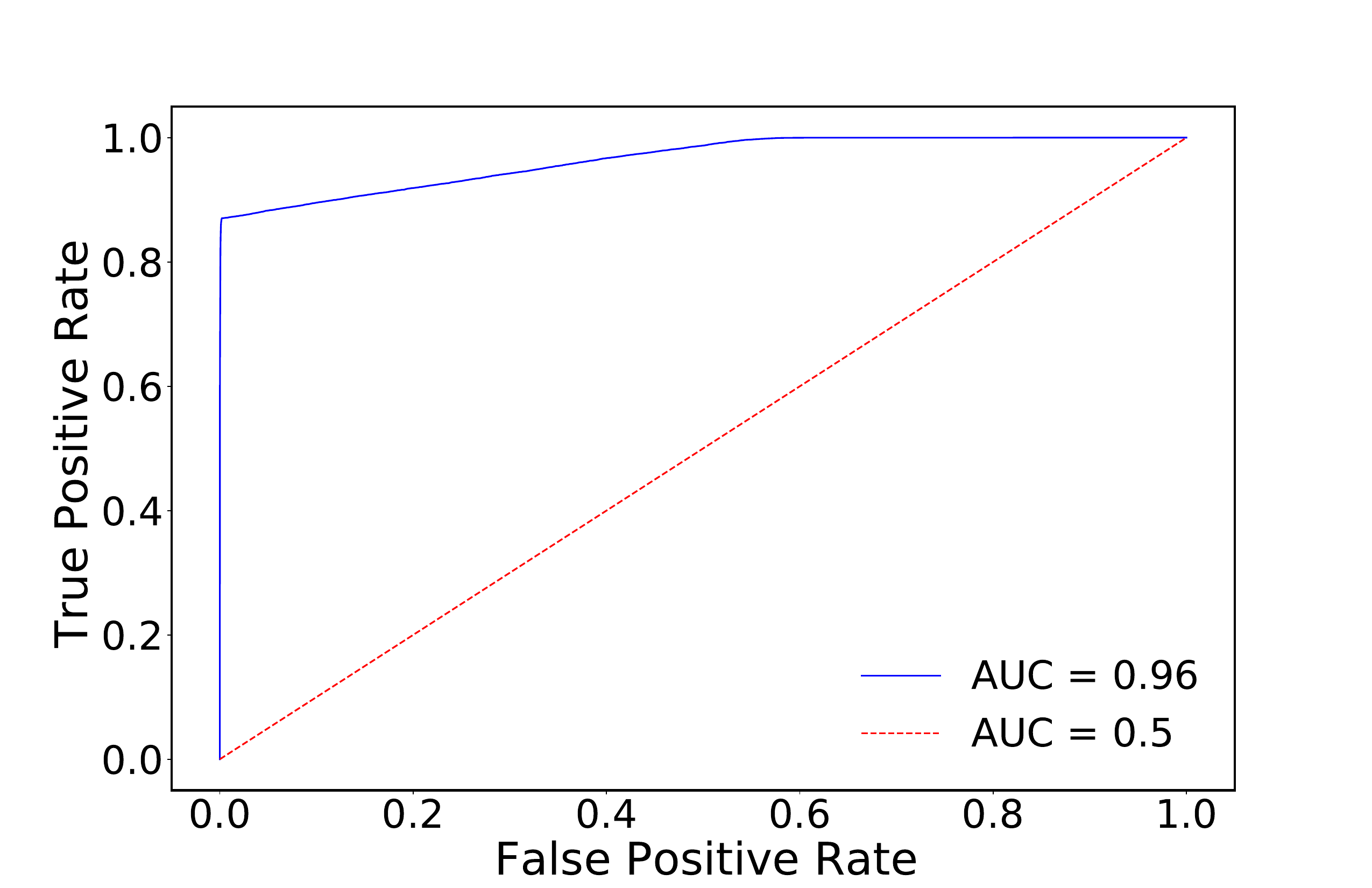}} 
   & \subfloat[ROC for Quenched dataset]{\hspace{-2.0em}\includegraphics[scale=0.18]{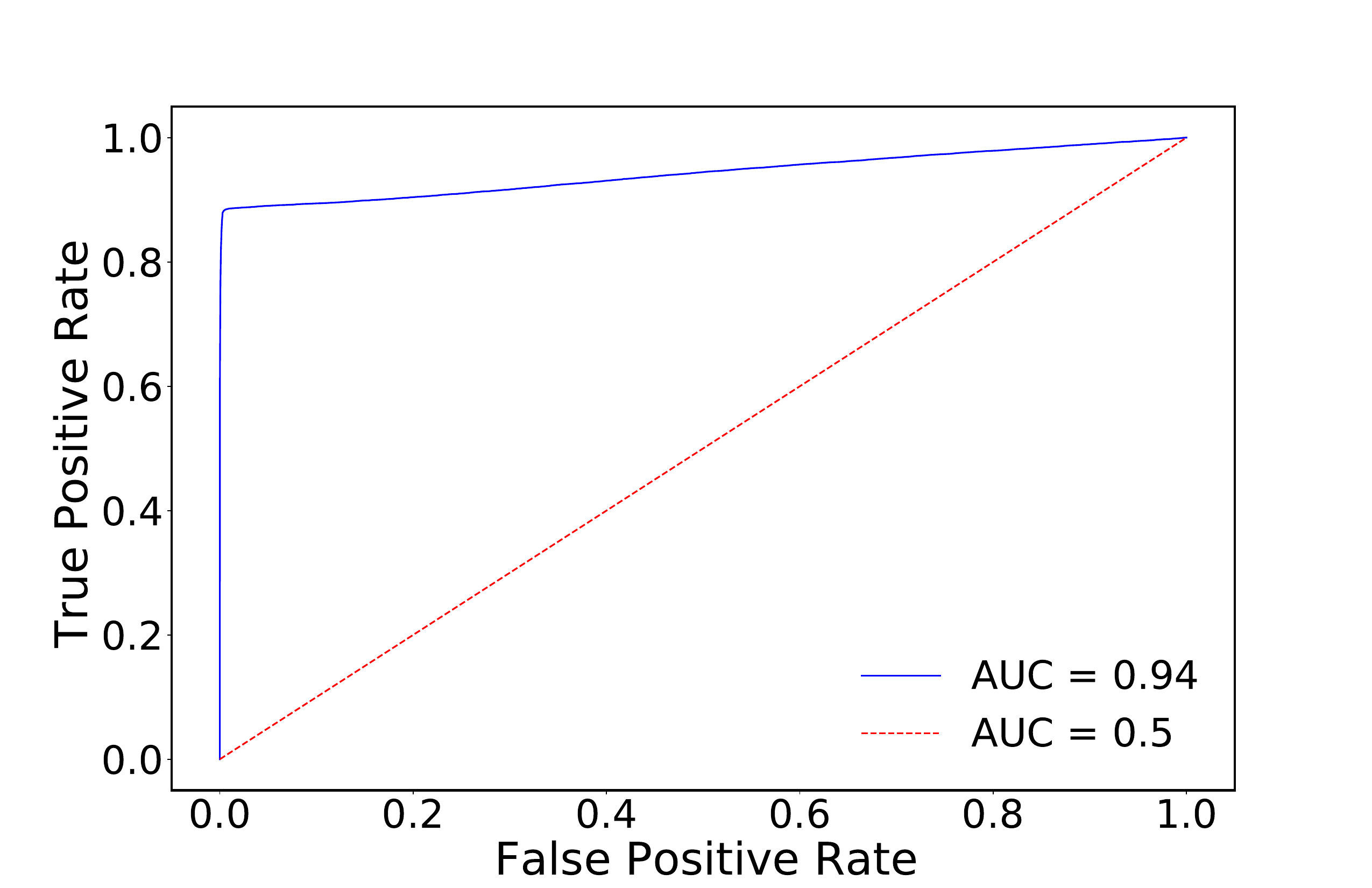}}\\
\end{tabular}
\end{tabularx}
\caption{(color online) ROC curves for both dataset of GraphRed model giving AUC (Area Under the Curve) parameter. The blue line (solid) represent the GraphRed model where as the red line (dotted) represents the general case when a model is not able to distinguish between positive and negative class.}\label{fig:roc}
\end{figure}

For both the datasets, we achieve accuracy around 98-99$\%$ in finding soft and hard type particles. The detailed training and validation accuracy w.r.t epochs for quenched and non-quenched datasets are available in Appendix A. Our model is evaluated using a diagnostic curve called Receiver Operating Characteristic (ROC) curve. ROC curve is a graphical plot that illustrates the diagnostic ability of a binary classification model through a parameter called the Area Under Curve (AUC) parameter. The AUC measures the ability of a classifier to distinguish between true and fake classes. Therefore, by plotting the ROC curve for both quenched and non-quenched dataset shown in Fig.~\ref{fig:roc}, we evaluate the performance of our GraphRed model to distinguish between hard and soft particles as two different classes. The AUC score of 0.5 indicates that the model predicts random classes and hence, is unable to discriminate between the classes, whereas the closeness of the AUC score to 1 represents the better ability of the model to distinguish between different classes. In the case of our dataset, we evaluated the AUC score of $0.94$ and $0.96$ for the quenched and non-quenched dataset, which can be inferred as the better performance of the model in finding hard and soft particles.

The output of GrapRed is the probability describing the particle as hard or soft, which is shown in Appendix A.1. Hence, it is essential to find a threshold value of probability such that if the probability is greater than the threshold value, then it is termed as a hard particle and vice-versa. In Table~\ref{tab:q_wp},\ref{tab:nq_wp} we have described the performance of various threshold values based on percentage of false negatives (FN) and false positives (FP).

\begin{table}[!hbt]
\centering
\begin{tabular}{|c|c|c|}
\hline
Threshold value& FP ($\%$)&FN ($\%$)\\
\hline 
0.5 & 0.07459 & 0.1695\\
0.7 & 0.06321 & 0.1873 \\
0.9 & 0.04612 & 0.2294 \\
\hline
\end{tabular}
\caption{\label{tab:q_wp} Threshold values of GraphRed model for non-quenched dataset}
\end{table}

\begin{table}[!hbt]
\centering
\begin{tabular}{|c|c|c|}
\hline
Threshold value & FP ($\%$)&FN ($\%$)\\
\hline 
0.5 & 0.1205 & 0.1798\\
0.7 & 0.0854 & 0.1717 \\
0.9 & 0.0094 & 0.4094 \\
\hline
\end{tabular}
\caption{\label{tab:nq_wp} Threshold values of GraphRed model for quenched dataset}
\end{table}

We observe that the FP ($\%$) is more for quenched datasets because quenching modifies various kinematic quantities, making hard particles more similar to soft particles. We chose the threshold values to reduce the miss-classifications by the cut value, which gives us the least FP ($\%$) and FN ($\%$). So, we decided the cut value of $0.7$ for both quenched and non-quenched datasets to keep a balance between FP ($\%$) and FN ($\%$). However, it is worth noting that in a heavy-ion collision, the number of particles observed from soft processes is much more in number than the hard processes. Accuracy is defined as the number of correctly identified particles by the total number of particles. Due to the high number of instances present, one class will lead the accuracy to a high value as the number of instances in the other class are low.  Therefore, accuracy must not be considered a sole measure of the goodness of the model. Hence, we have performed additional physics validation for the predicted hard particles by our model to demonstrate its effectiveness in the presence of a large background.

\subsubsection{Embedded event by PYTHIA $\&$ HYDJET}
We train the GraphRed model with a million events of PYTHIA embedded with HYDJET background. The model is trained by embedding the final state particles of each event into its geometrical embedding. The training procedure deployed is the same as described in Sec~\ref{sec:hydjet_train}. The training and validation accuracy plots w.r.t epochs are given in Appendix A.2.
\begin{figure}[!hbt]
    \centering
    \includegraphics[scale=0.2]{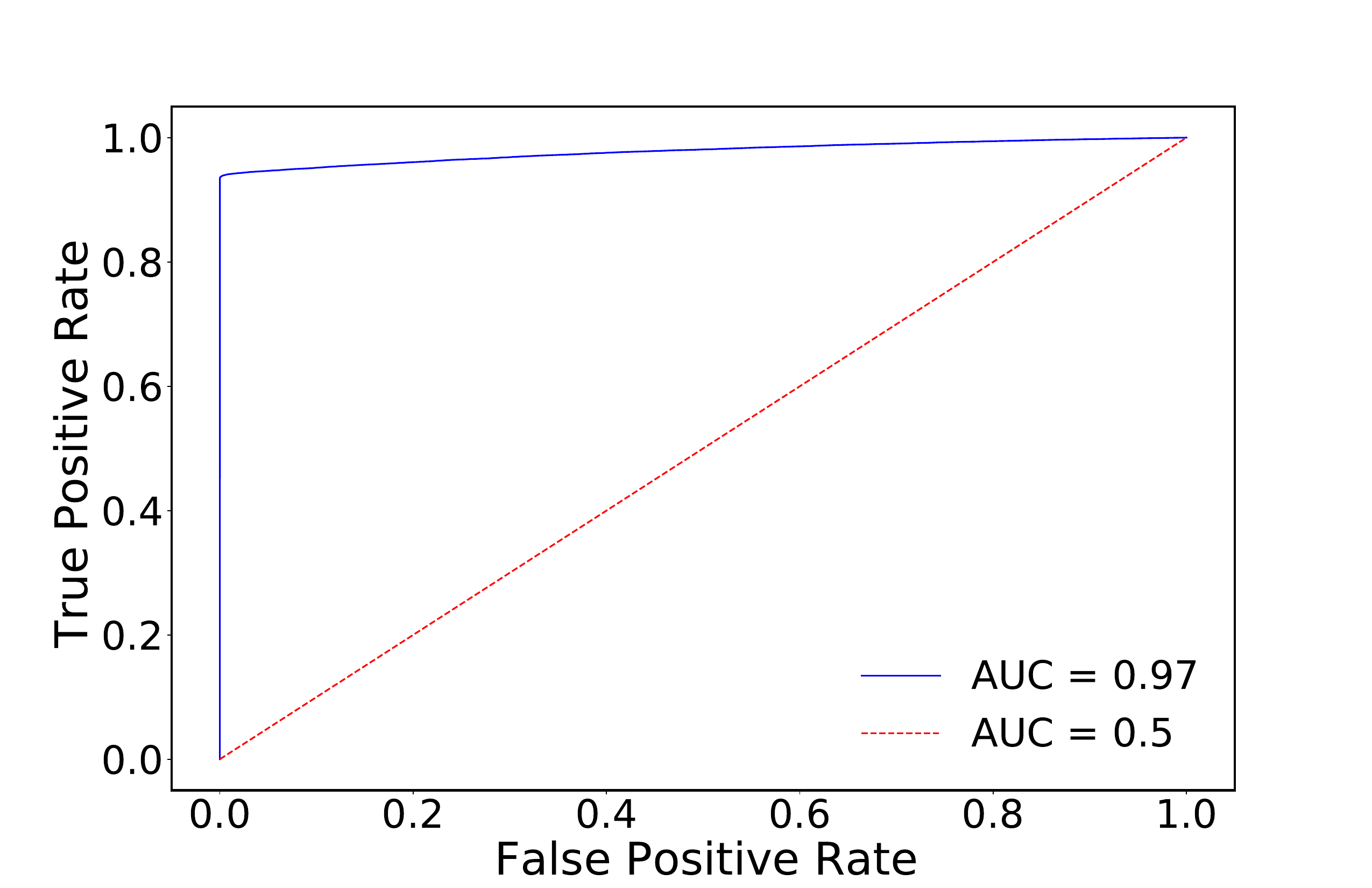}
    \caption{(color online) ROC curve for embedded event dataset giving AUC parameter. The blue line (solid) represent the GraphRed model where as the red line (dotted) represents the general case when a model is not able to distinguish between positive and negative class.}
    \label{fig:roc2}
\end{figure}

We observe that we achieved an accuracy of $>99\%$ in characterizing particles into hard and soft. We evaluated our model in terms of the ROC curve.  The ROC curve is shown in Fig.~\ref{fig:roc2}. In the case of the embedded event dataset, we found the AUC score of $\sim 0.97$, which describes a good performance of the model to discriminate between hard and soft particles. Moreover, similar to the previous case, we have analyzed the performance of various threshold values based on the percentage of false negatives (FN) and false positives (FP), as shown below. The output probability distribution by GraphRed is also provided in Appendix A.2 . 

\begin{table}[!hbt]
\centering
\begin{tabular}{|c|c|c|}
\hline
Threshold value & FP ($\%$)&FN ($\%$)\\
\hline 
0.5 & 0.0032 & 0.1244\\
0.7 & 0.0027 & 0.1252 \\
0.9 & 0.0022 & 0.1261 \\
\hline
\end{tabular}
\caption{\label{tab:nq_pythia} Threshold values of GraphRed model for embedded event dataset}
\end{table}

Moreover,from Table.~\ref{tab:nq_pythia} we see that there is not much change in FP ($\%$) and FN ($\%$) with respect to the threshold value. Here, also we chose the cut value of $0.7$ to keep a balance between the  FP ($\%$) and FN ($\%$) and provided further analyses with particle and jet-level variables in the above sections.

\subsection{Particle-level comparison}
\begin{figure}[!hbt]
\def\tabularxcolumn#1{m{#1}}
\begin{tabularx}{\linewidth}{@{}cXX@{}}
\begin{tabular}{cc}
\subfloat[$p_{T}$ Distribution]{\hspace{-1.0em}\includegraphics[scale=0.4]{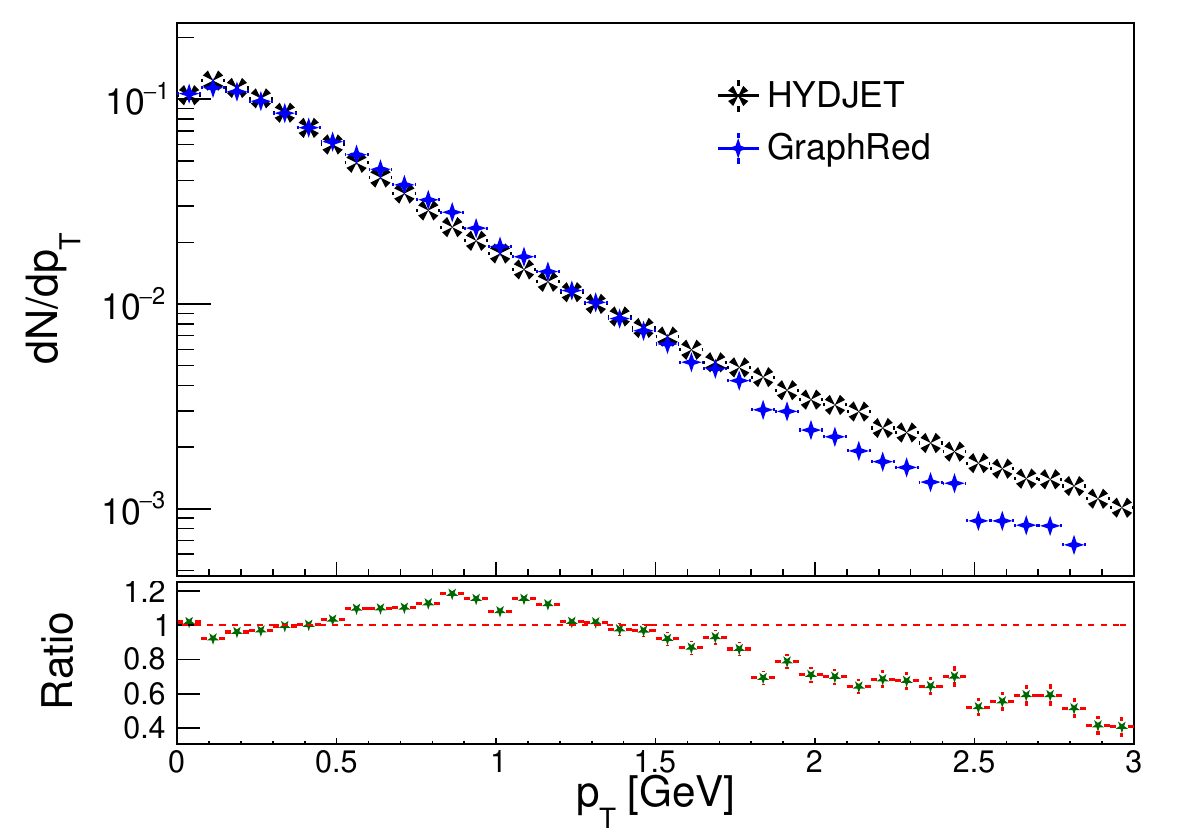}} 
   & \subfloat[Energy Distribution]{\hspace{-1.0em}\includegraphics[scale=0.4]{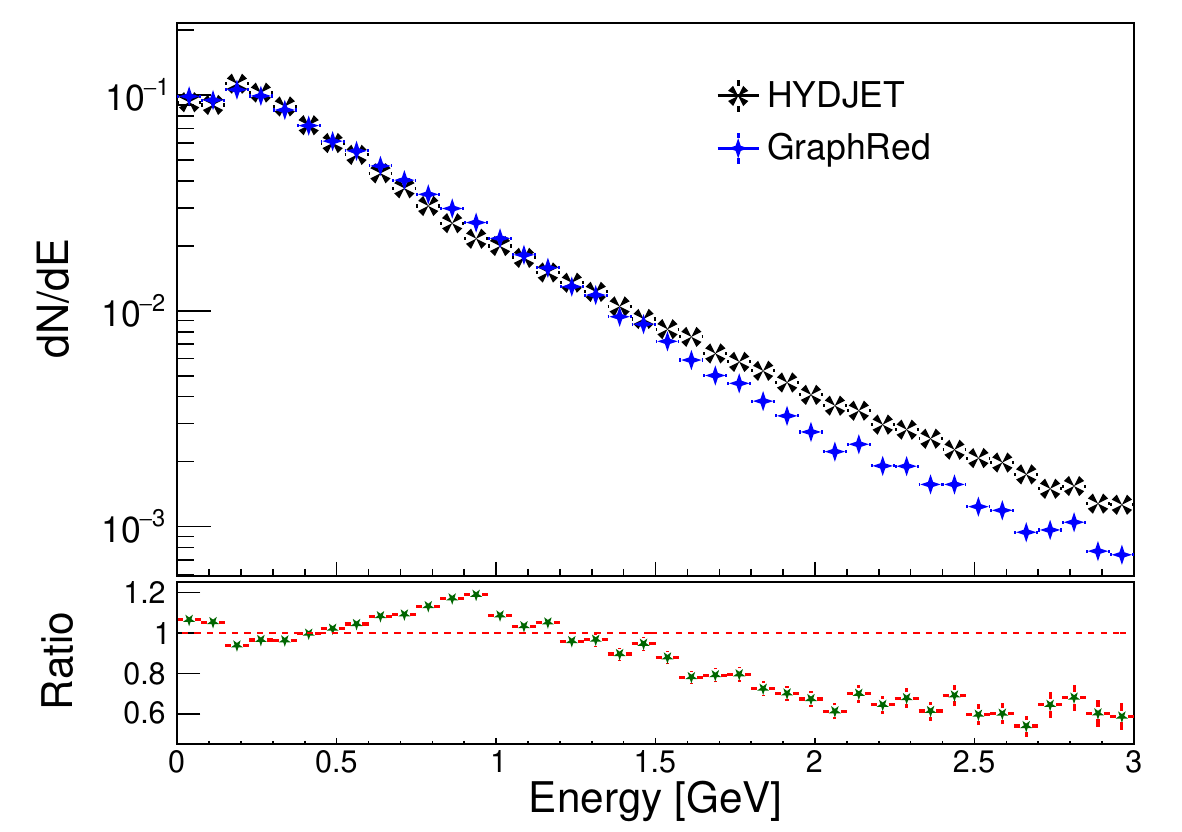}}\\
\end{tabular}
\end{tabularx}
\caption{(color online) Distribution comparison between HYDJET++ (data) hard particles and GraphRed (reco) most likely non-quenched hard particles}\label{fig:comp}
\end{figure}

\begin{figure}[!hbt]
\def\tabularxcolumn#1{m{#1}}
\begin{tabularx}{\linewidth}{@{}cXX@{}}
\begin{tabular}{cc}
\subfloat[$p_{T}$ Distribution]{\hspace{-1.0em}\includegraphics[scale=0.39]{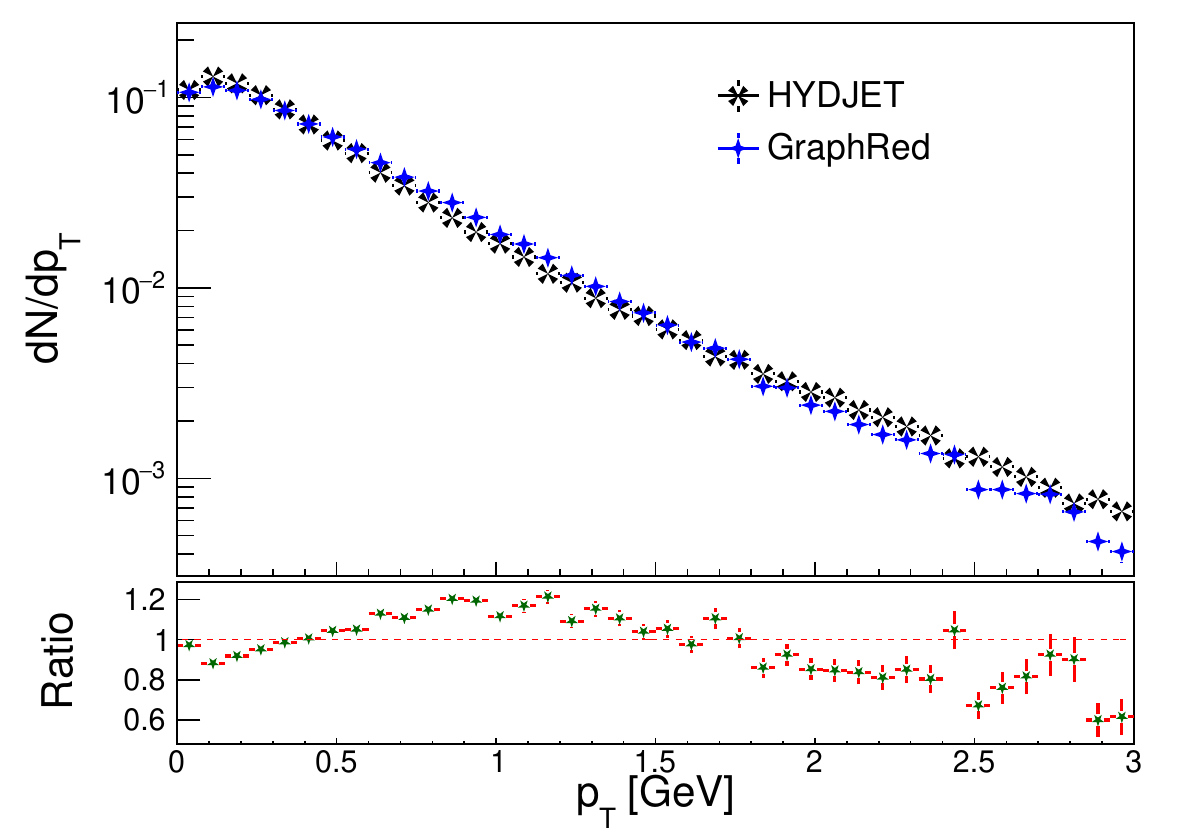}} 
   & \subfloat[Energy Distribution]{\hspace{-1.0em}\includegraphics[scale=0.39]{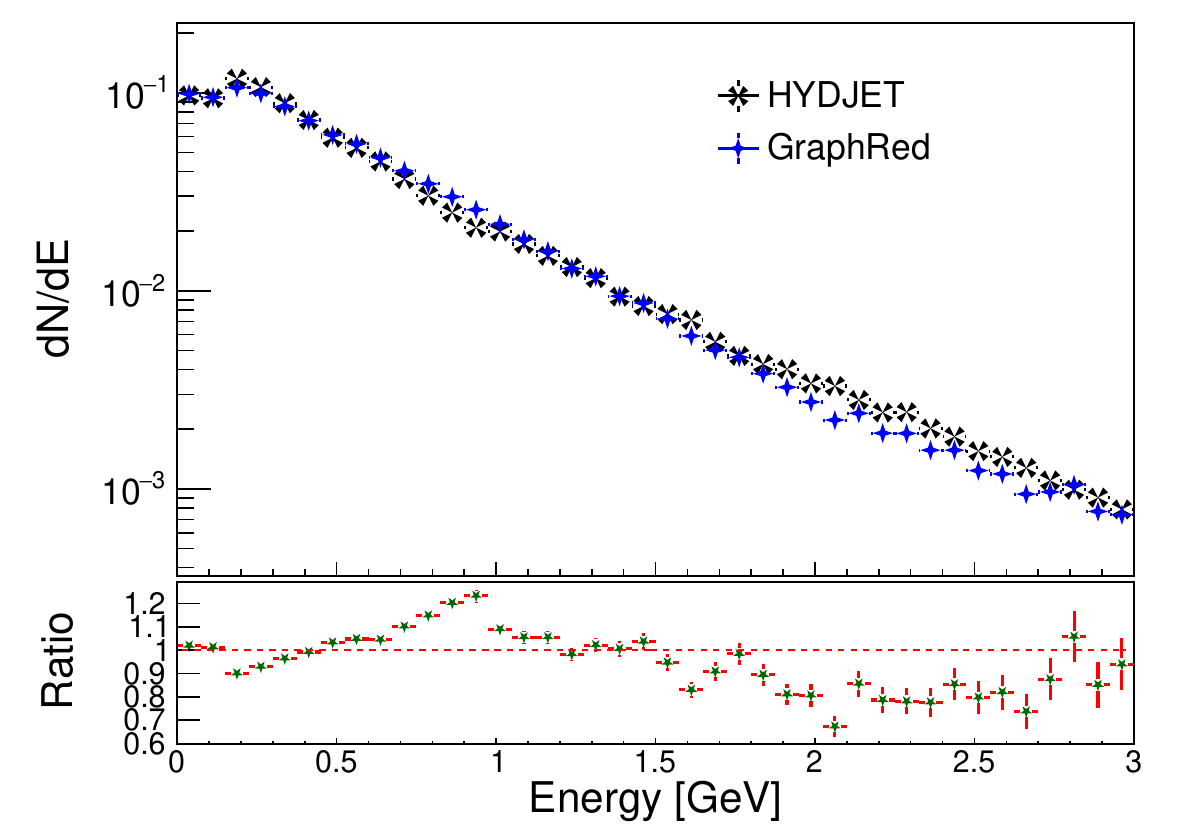}}\\
\end{tabular}
\end{tabularx}
\caption{(color online) Distribution comparison between HYDJET++ (data) hard particles and GraphRed (reco) most likely quenched hard particles}\label{fig:comp_quench}
\end{figure}

\begin{figure}[!hbt]
\def\tabularxcolumn#1{m{#1}}
\begin{tabularx}{\linewidth}{@{}cXX@{}}
\begin{tabular}{cc}
\subfloat[$p_{T}$ Distribution]{\hspace{-1.0em}\includegraphics[scale=0.39]{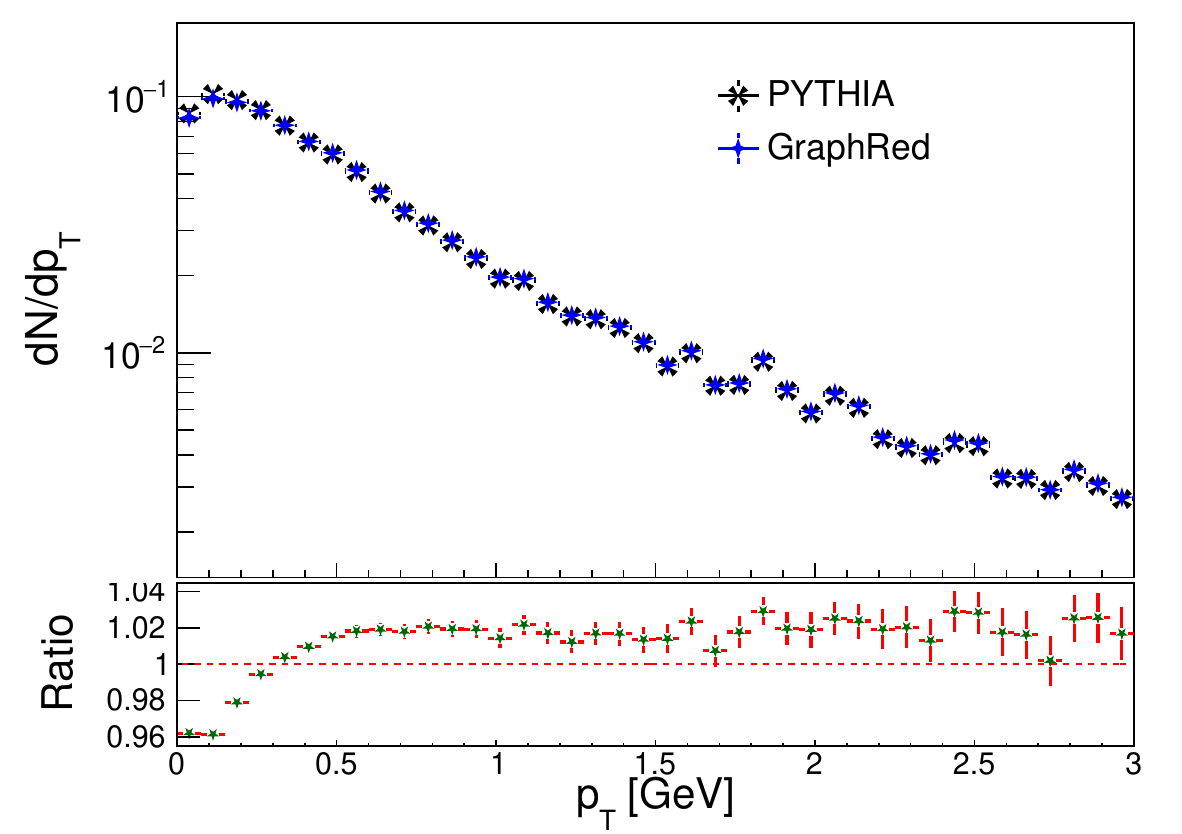}} 
   & \subfloat[Energy Distribution]{\hspace{-1.0em}\includegraphics[scale=0.39]{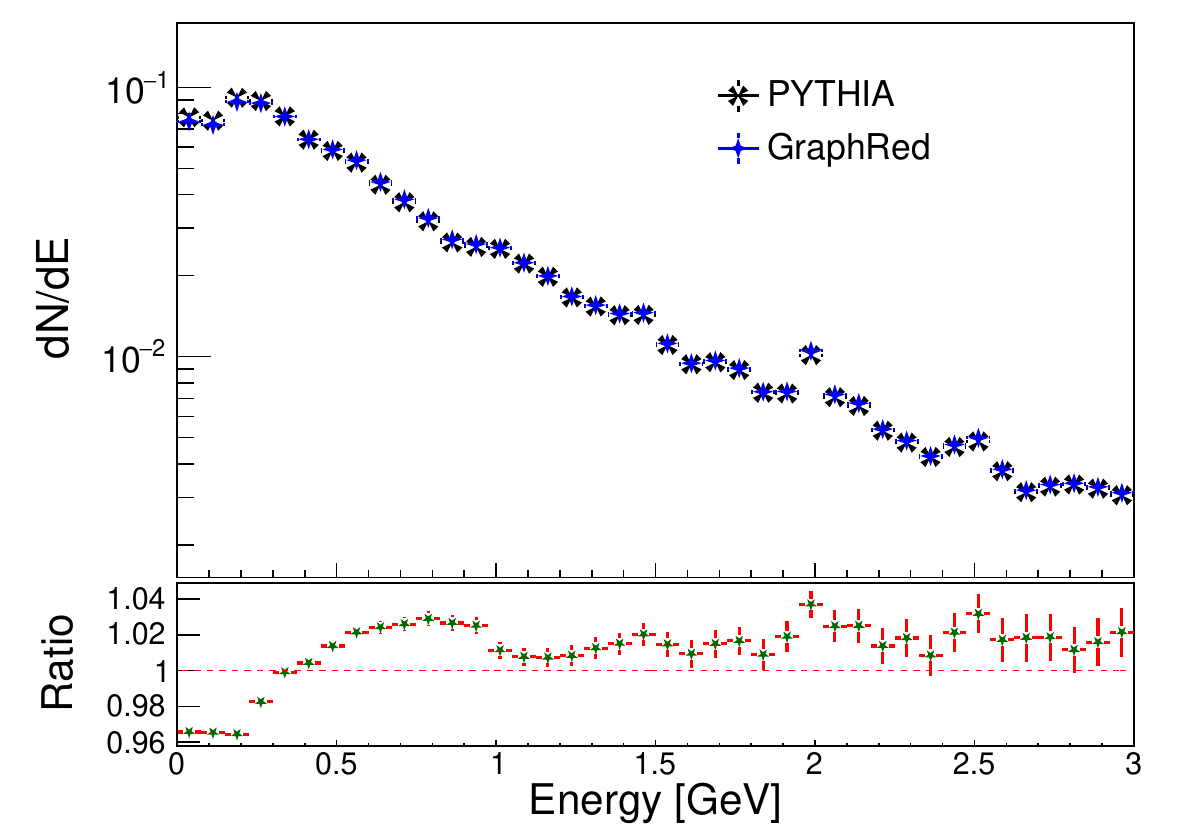}}\\
\end{tabular}
\end{tabularx}
\caption{(color online) Distribution comparison between PYTHIA (data) hard particles and GraphRed (reco) most likely hard particles}\label{fig:comp_pythia}
\end{figure}

As we discussed in the above section, accuracy is not the sole measure of the goodness of the model. We extend the evaluation of the model by comparison of various physical variables on particle-level like $p_{T}$ spectra, energy, etc. We compared the particle-level physical observable distribution obtained from simulations (full heavy-ion event by HYDJET and embedded event by HYDJET $\&$ PYTHIA) serving as truth and our GraphRed (reconstructed) method. The method is evaluated for both full and embedded event datasets. For full heavy-ion event by HYDJET, the $p_{T}$ and energy distribution for quenched data-set is shown in Fig.~\ref{fig:comp} and for non-quenched dataset in Fig.~\ref{fig:comp_quench}. A ratio plot of data vs reconstructed is also shown, describing the effectiveness of the model. Fig.~\ref{fig:comp_pythia} describes the $p_{T}$ and energy distribution for the embedded events by PYTHIA $\&$ HYDJET with ratio plot of data vs reconstructed shown below.

We infer from the plots that GraphRed is able to reconstruct the $p_{T}$ and energy distribution to a good extent which demonstrated the effectiveness in finding the most likely hard particles in an event. Moreover, it is worth noting that our proposed method is very much able to find the low $p_{T}$ hard particles as the ratio hovers around unity for the low $p_{T}$ range for the datasets.
\newpage
\subsection{Jet-level comparison}
\label{sec:fastjet}

To demonstrate the effectiveness of our proposed method to find hard particles, we
compared our technique with some of the present benchmark techniques: (a) background subtraction and reconstruction methods and (b) performance of background reduction methods.

\begin{figure}[!hbt]
\def\tabularxcolumn#1{m{#1}}
\begin{tabularx}{\linewidth}{@{}cXX@{}}
\begin{tabular}{cc}
\subfloat[$\rho$ parameter]{\hspace{-1.0em}\includegraphics[scale=0.4]{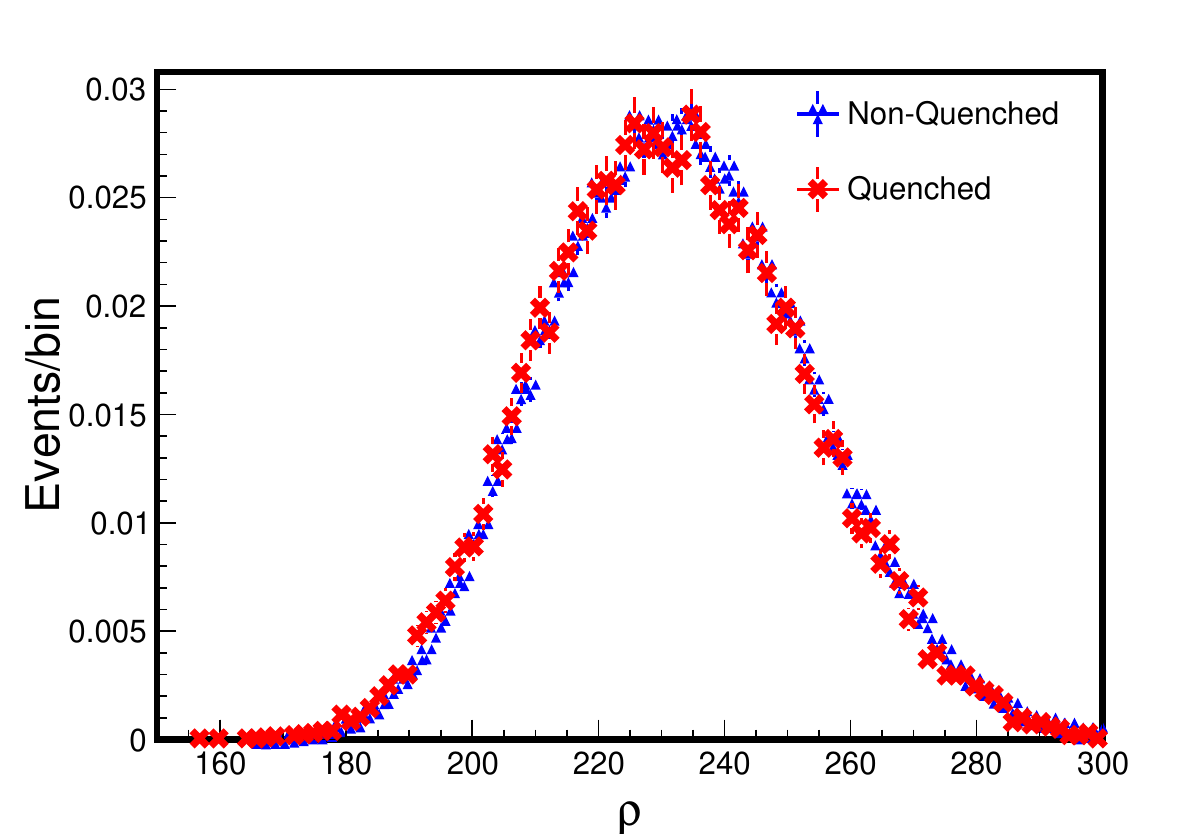}} 
   & \subfloat[$\sigma$ parameter]{\hspace{-1.0em}\includegraphics[scale=0.4]{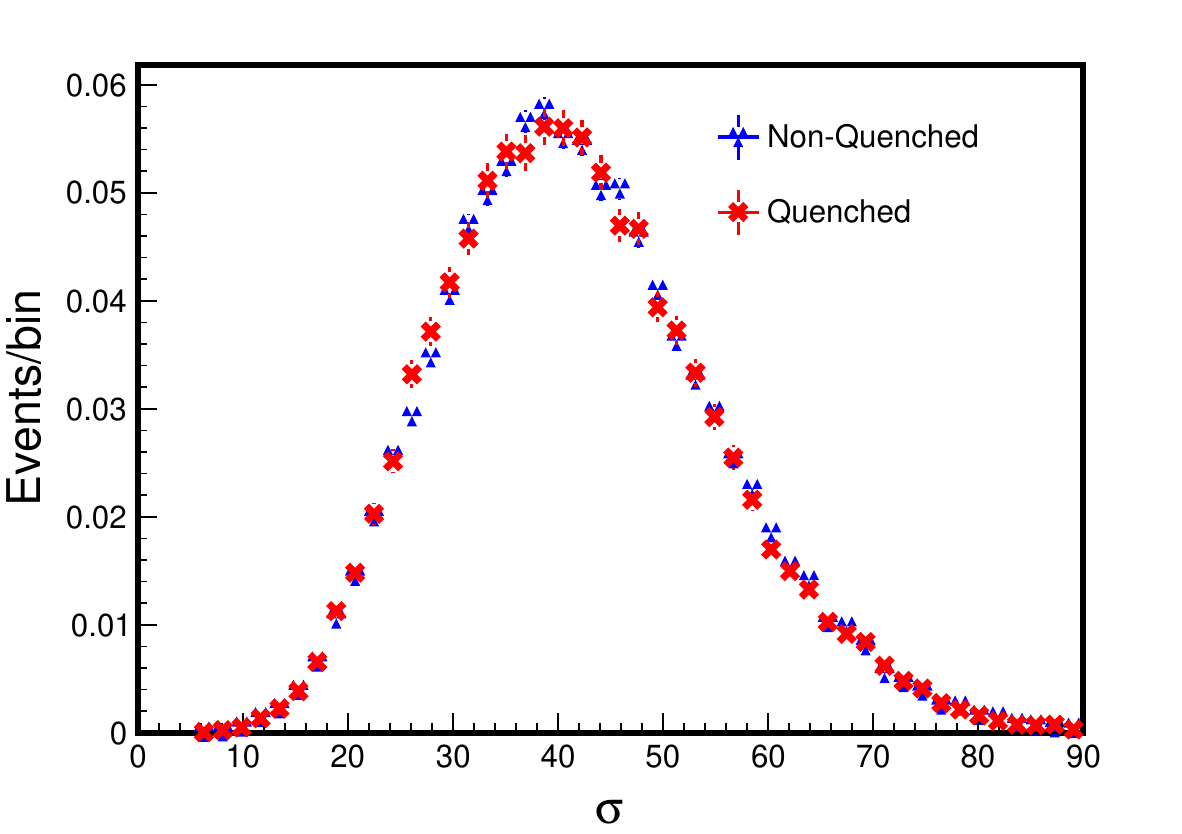}}\\
\end{tabular}
\end{tabularx}
\caption{(color online) $\rho$ [GeV/A] where A is the area of jet $\&$ $\sigma$ ( estimated in the $p_{T}$ density per unit $\sqrt{A}$) parameter comparison for quenched and non-quenched data-sets by excluding the two-hardest jets while modelling}\label{fig:rho}
\end{figure}

\begin{figure}[!hbt]
\def\tabularxcolumn#1{m{#1}}
\begin{tabularx}{\linewidth}{@{}cXX@{}}
\begin{tabular}{cc}
\subfloat[$\rho$ parameter]{\hspace{-1.0em}\includegraphics[scale=0.4]{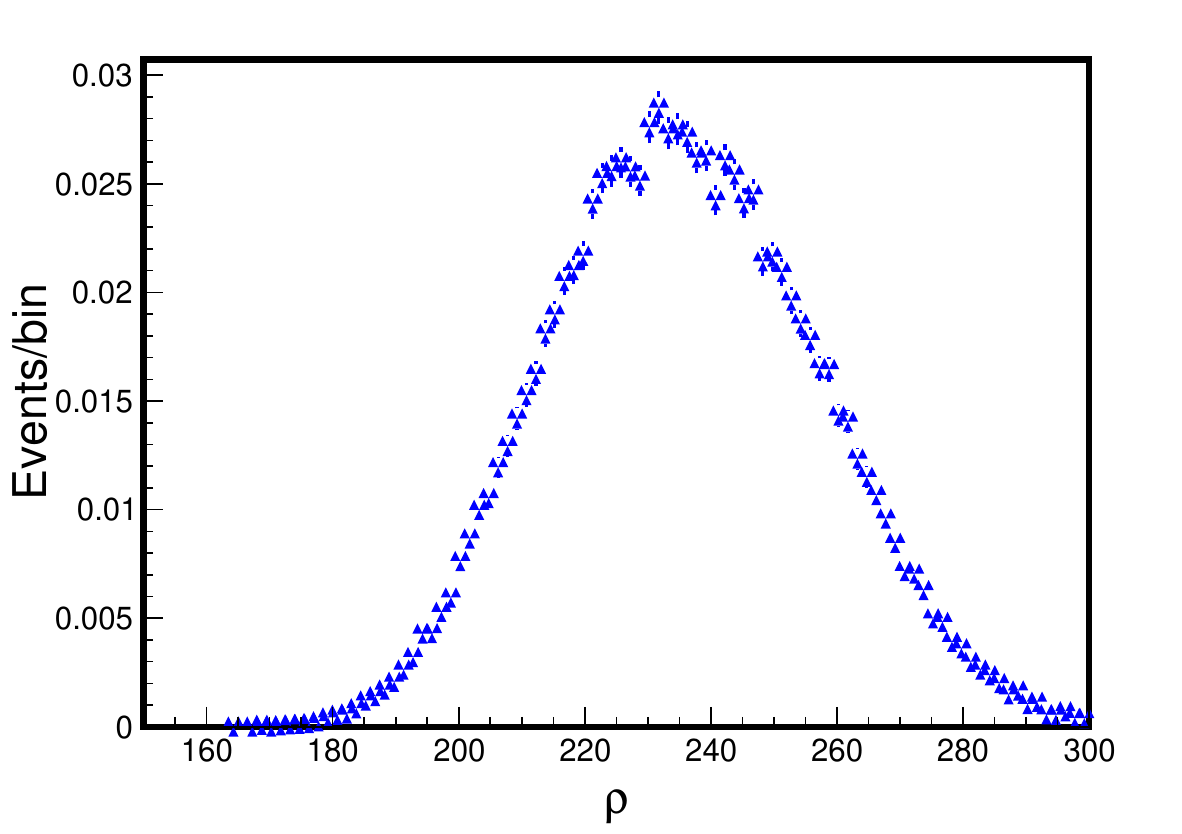}} 
   & \subfloat[$\sigma$ parameter]{\hspace{-1.0em}\includegraphics[scale=0.4]{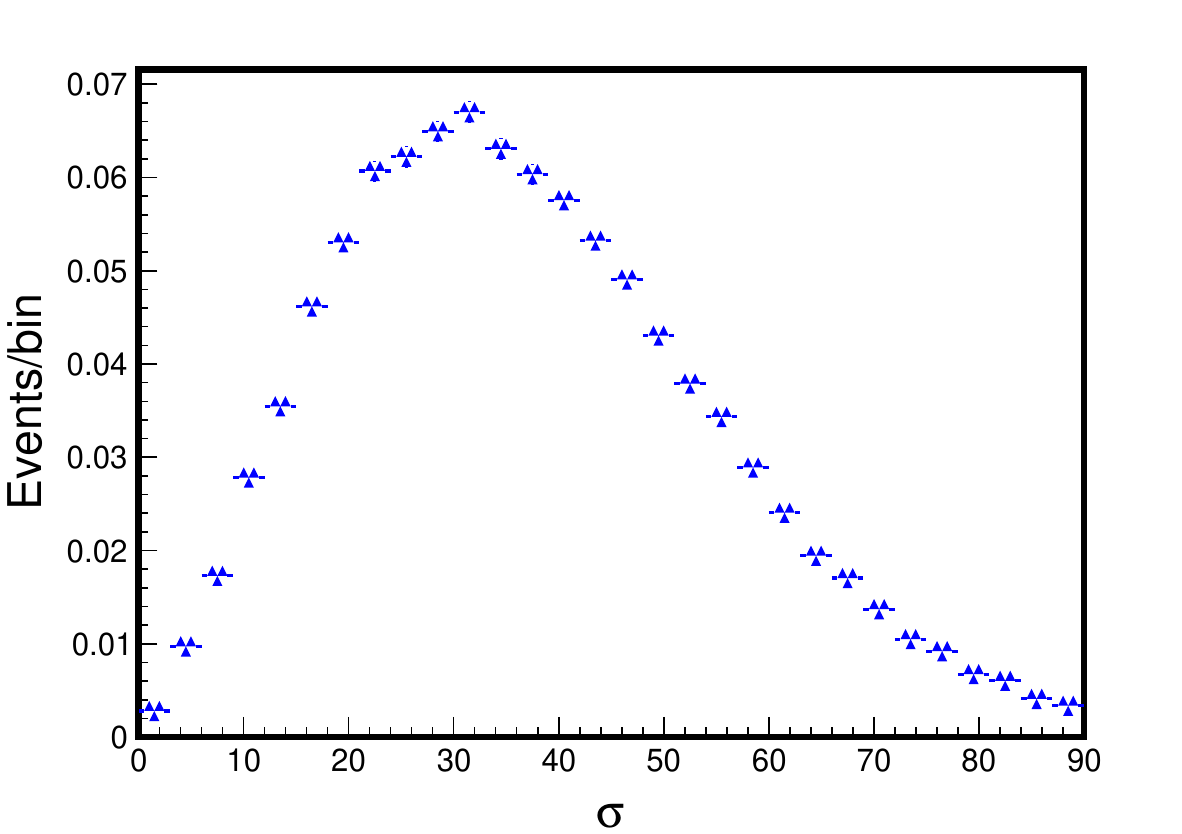}}\\
\end{tabular}
\end{tabularx}
\caption{(color online) $\rho$ [GeV/A] where A is the area of jet $\&$ $\sigma$ ( estimated in the $p_{T}$ density per unit $\sqrt{A}$) parameter for embedded event dataset}\label{fig:rho_pythia}
\end{figure}

The first technique we implemented is fastjet~\cite{fastjet} area-based reduction. This method removes the background contamination from jets by estimating the background and subtracting its contamination from the transverse momentum ($p_{T}$) of jets. We implemented an area-based reduction method present in fastjet that estimates background by taking jet definition, area definition, and selector. The jet algorithm used and its parameter constitute the jet definition. In our case, we used anti-$k_{T}$ algorithm with $R = 0.6$ ($R$ is the radius of jet) where jets having a $p^{min}_{T}>8$ $GeV$ are considered . The area definition provides information about the jet areas, which measure the surface over which a jet extends and gives us a measure of a jet’s susceptibility to soft contamination. The input is the area type and a maximum rapidity range the specification of soft massless (ghost) particles. We used the following parameters for the analysis i.e., the area type, maximum rapidity $y = 1$ with a grid spacing of $0.4$ and excluded the two-hardest jets while modelling background. This method produces two output parameters $\rho$ ($p_{T}$ density per unit area A near jet) $\&$ $\sigma$ ( an estimate of the fluctuations in the $p_{T}$ density per unit $\sqrt{A}$) for background and then utilize it to subtract its contamination given below and proposed in Ref.~\cite{Cacciari_2008}. 
\begin{equation}
    p_{T,jet}^{sub} = p_{T,jet}^{raw} - \rho\cdot A_{jet}
\end{equation}
The above technique was applied to all particles of the event. The plot of background estimation parameters $\rho$ $\&$ $\sigma$ is shown in Fig.~\ref{fig:rho} and Fig.~\ref{fig:rho_pythia}. 

The second technique that we checked to compare our GraphRed approach is SoftKiller~\cite{softkiller}. This method involves eliminating particles below some $p_{T}$ threshold, $p^{thres}_{T}$ over a patch, such that $\omega$ is zero. $\omega$ is an event-wide estimate of transverse momentum flow density in the area–median approach~\cite{five,six}.  The event is first broken into small patches, and the $\omega$ is computed as the median of the transverse momentum flow density per unit area in rapidity-azimuth across all the patches. 
\begin{equation}
    \omega = \underset{\substack{i\in patches}}{median} (\frac{p_{ti}}{A_{ti}})
\end{equation}
where $p_{ti}$ and $A_{ti}$ are  the transverse momentum and area of $i^{th}$ patch  respectively. The event is initially clustered by anti-$k_{T}$ algorithm with $R = 0.6$ using the particle of the entire event, and then Softkiller is applied. The  Softkiller operation is performed with a grid size of 0.3. And as a next and final step, the particles are again clustered using anti $k_{T}$ algorithm with $R = 0.6$ where jets are having a $p^{min}_{T}>8$ $GeV$ are considered.


The third and last technique implemented is event-wide constituent subtractor (CS) ~\cite{CS,It_CS}. The basis of this method is almost similar to the area-based reduction method. The ghost particles are used, but the constituent-level subtraction is performed at the particle level, which leads to a simultaneous correction in the jet as well as its substructure. This is done by combining the kinematics of constituent particles of a jet with the kinematics of soft particles that are added to balance the pileup contribution. All particles of an event are clustered using the anti $k_{T}$ algorithm with $R = 0.6$, and then event-wide CS is applied to obtain the suitable value of parameters. 
A default background estimator is used with the grid size $(0.3)$. After the application of event wide CS, corrected particles are again clustered using anti $k_{T}$ algorithm with $R = 0.6$ where where jets having a $p^{min}_{T}>8$ $GeV$ are considered. \\
The performance of the methods mentioned above is evaluated against the truth and our GraphRed strategy when serving as a pre-processing method. The hard particles in an event predicted by GraphRed using the threshold as mentioned earlier value are clustered using anti-$k_{T}$ algorithm with $R = 0.6$. The true distributions are obtained by using the HYDJET information, and the clustering is done by using the anti-$k_{T}$ algorithm with $R = 0.6$. In both cases, the transverse momentum threshold of jets are uniformly taken as $p^{min}_{T}>8$ $GeV$.

\begin{figure}[!hbt]

\begin{minipage}{.32\linewidth}
\centering
\subfloat[$p_{T}$ weighted $(\eta, \phi)$ plane of entire event]{\label{main:a2}\includegraphics[scale=.25]{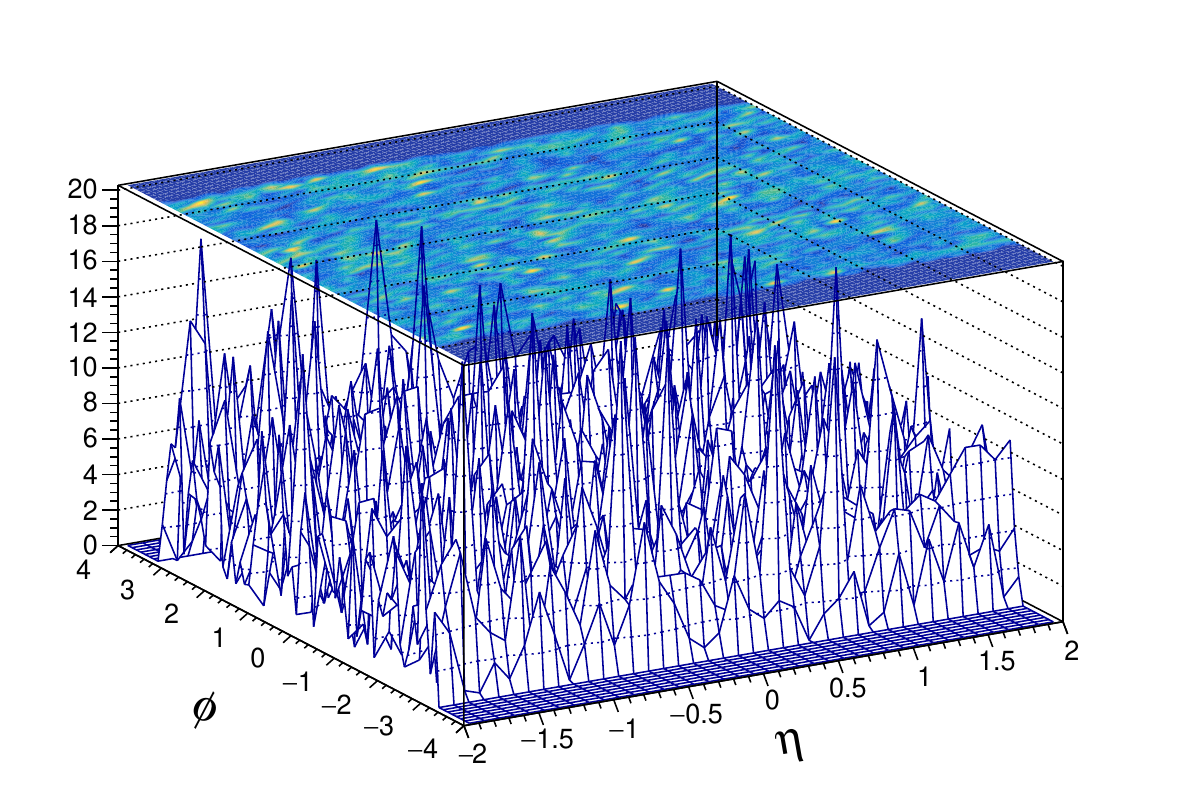}}
\end{minipage}%
\begin{minipage}{.40\linewidth}
\centering
\subfloat[$p_{T}$ weighted $(\eta, \phi)$ plane of jet particles]{\label{main:a2}\includegraphics[scale=.25]{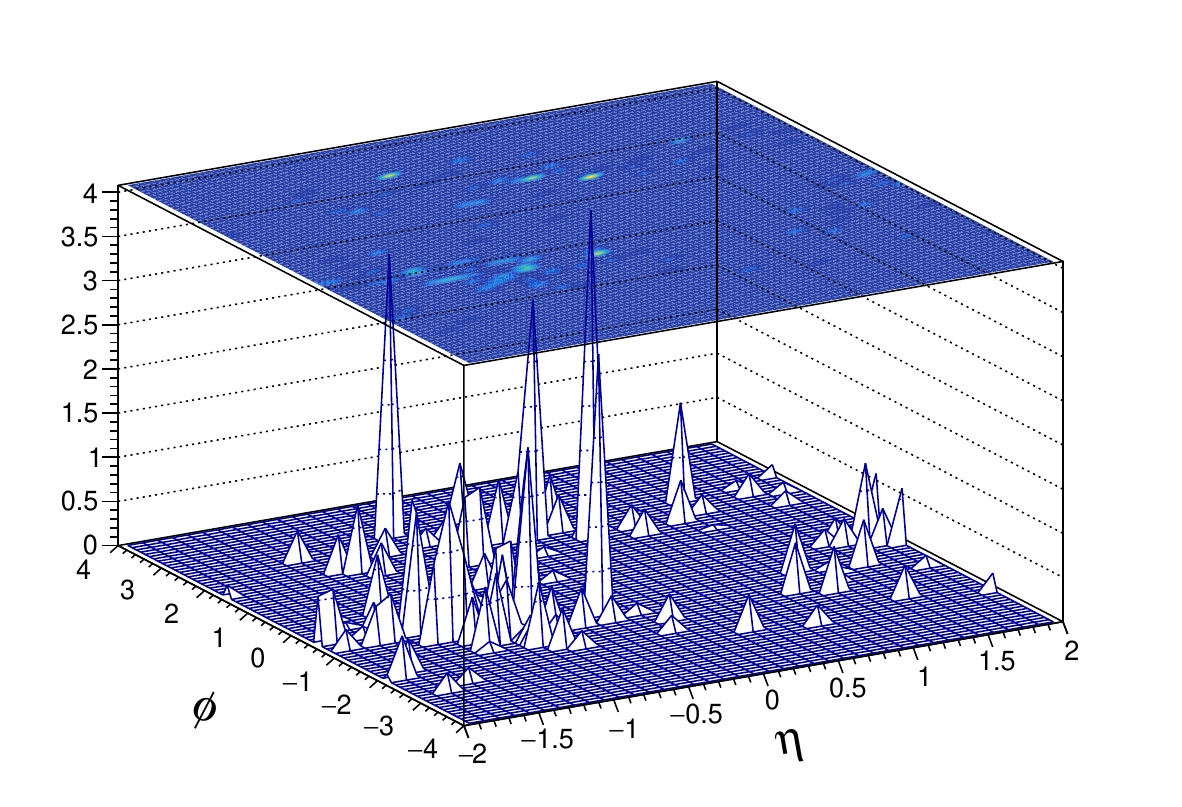}}
\end{minipage}
\begin{minipage}{.25\linewidth}
\centering
\subfloat[$p_{T}$ weighted $(\eta, \phi)$ plane of leading jet particles]{\label{main:b2}\includegraphics[scale=.25]{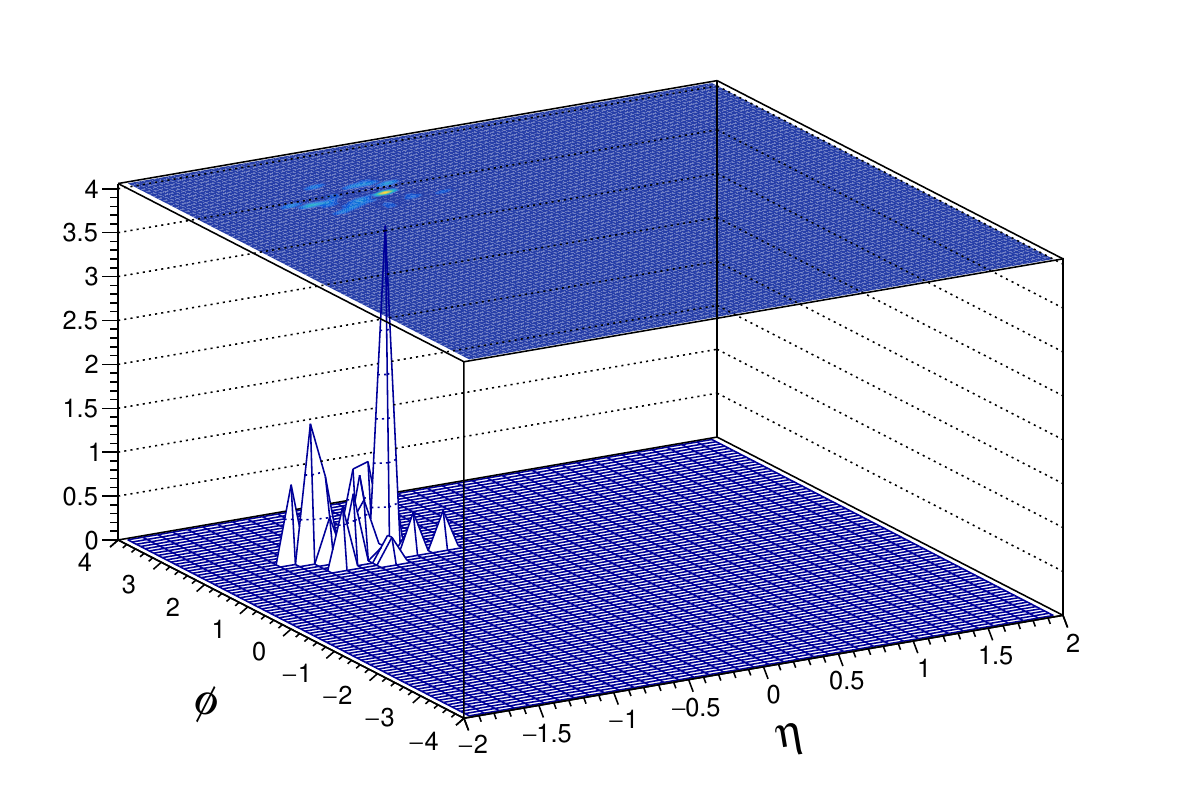}}
\end{minipage}\par\medskip

\caption{(color online) Representation of particles in an event in $(\eta, \phi)$ plane weighted with $p_T$ }
\label{fig:jet_show}
\end{figure}

\begin{figure}[!hbt]

\begin{minipage}{.5\linewidth}
\centering
\subfloat[$ p_{T}$ for leading jet]{\label{main:a}\includegraphics[scale=.4]{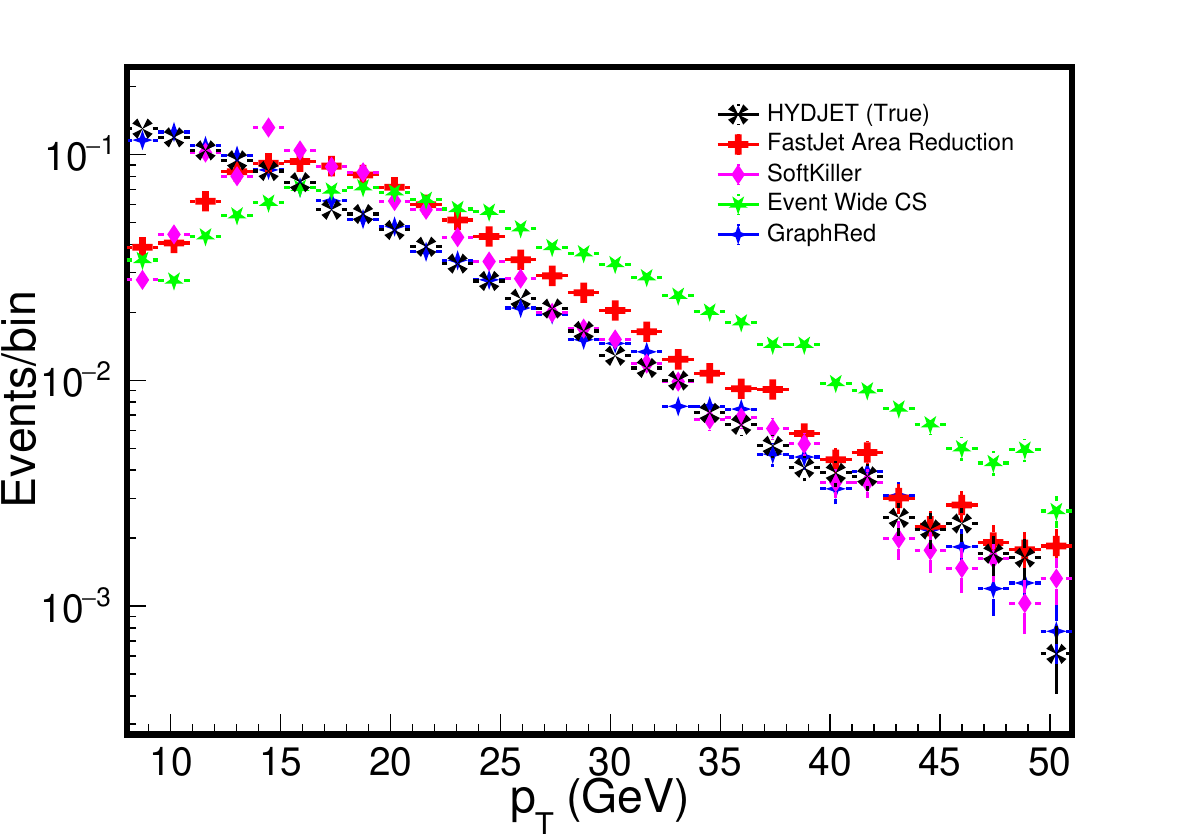}}
\end{minipage}%
\begin{minipage}{.5\linewidth}
\centering
\subfloat[$ E_{T}$ for leading jet]{\label{main:b}\includegraphics[scale=.4]{Dist_pt_noquench.pdf}}
\end{minipage}\par\medskip

\caption{(color online) $p_{T}$ and $ E_{T}$ distribution for leading jet in non-quenched dataset. The event compared with the background reduction techniques. The HYDJET (True) is the true distribution obtained by clustering of hard particles. GraphRed distribution is the distribution obtained by clustering the hard particles predicted by the model.}
\label{fig:ljet_nq}
\end{figure}

\begin{figure}

\begin{minipage}{.5\linewidth}
\centering
\subfloat[$ p_{T}$ for leading jet]{\label{main:a2}\includegraphics[scale=.4]{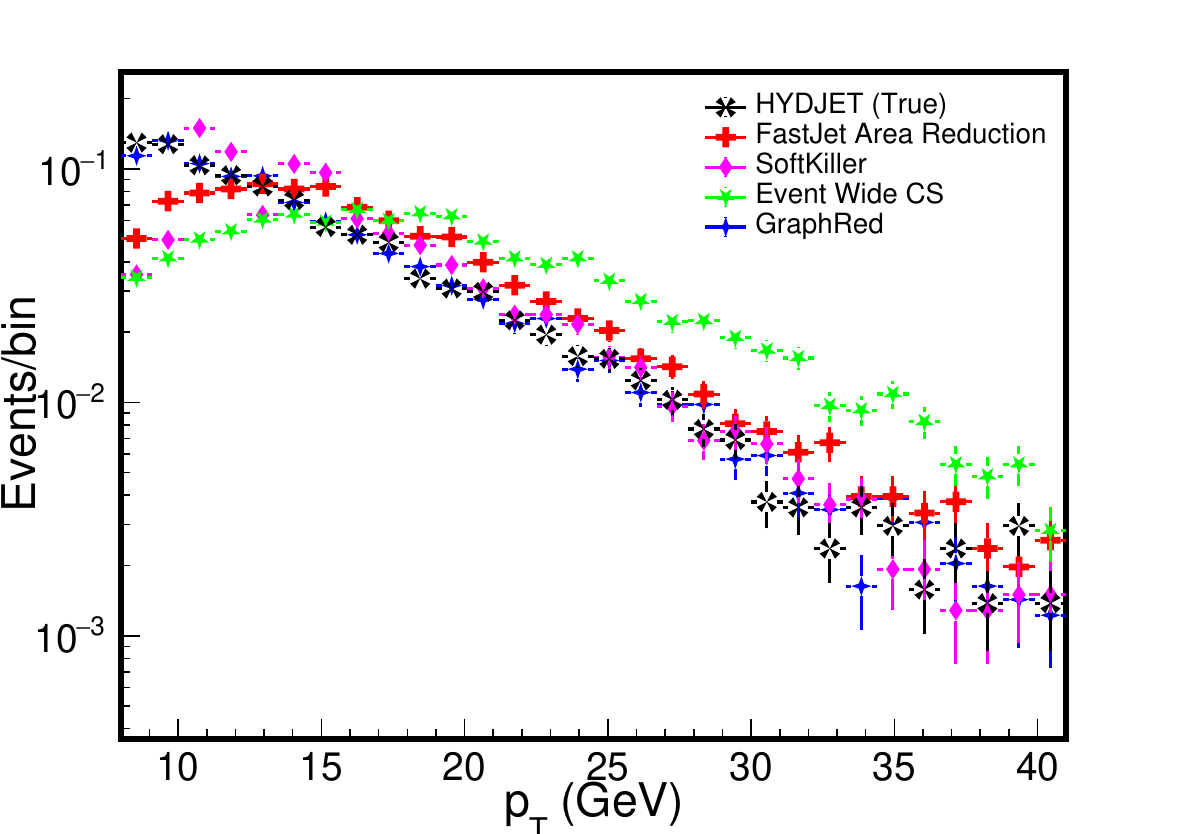}}
\end{minipage}%
\begin{minipage}{.5\linewidth}
\centering
\subfloat[$ E_{T}$ for leading jet]{\label{main:b2}\includegraphics[scale=.4]{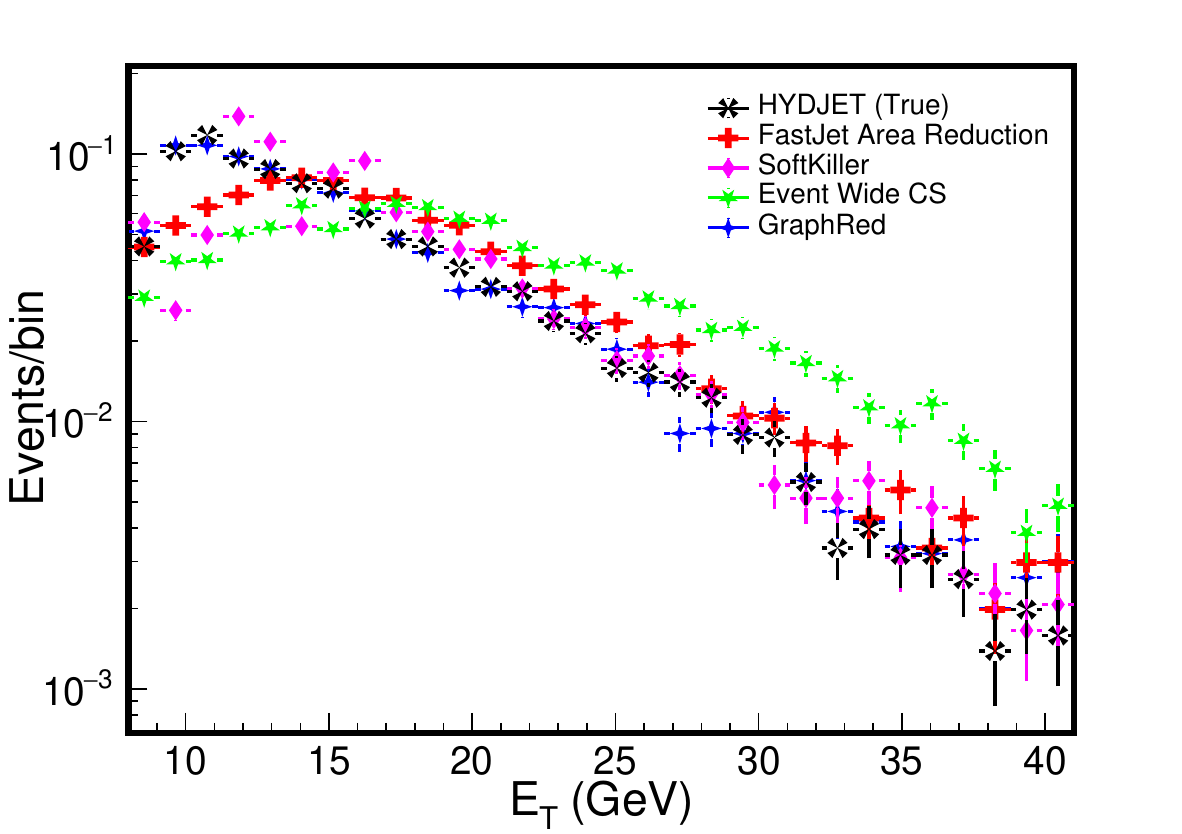}}
\end{minipage}\par\medskip

\caption{(color online) $p_{T}$ and $ E_{T}$ distribution for leading jet in quenched dataset. The event compared with the background reduction techniques. The HYDJET (true) is the true distribution obtained by clustering of hard particles. GraphRed distribution is the distribution obtained by clustering the hard particles predicted by the model.}
\label{fig:ljet_q}
\end{figure}

\begin{figure}

\begin{minipage}{.5\linewidth}
\centering
\subfloat[$ p_{T}$ for leading jet]{\label{main:a}\includegraphics[scale=.4]{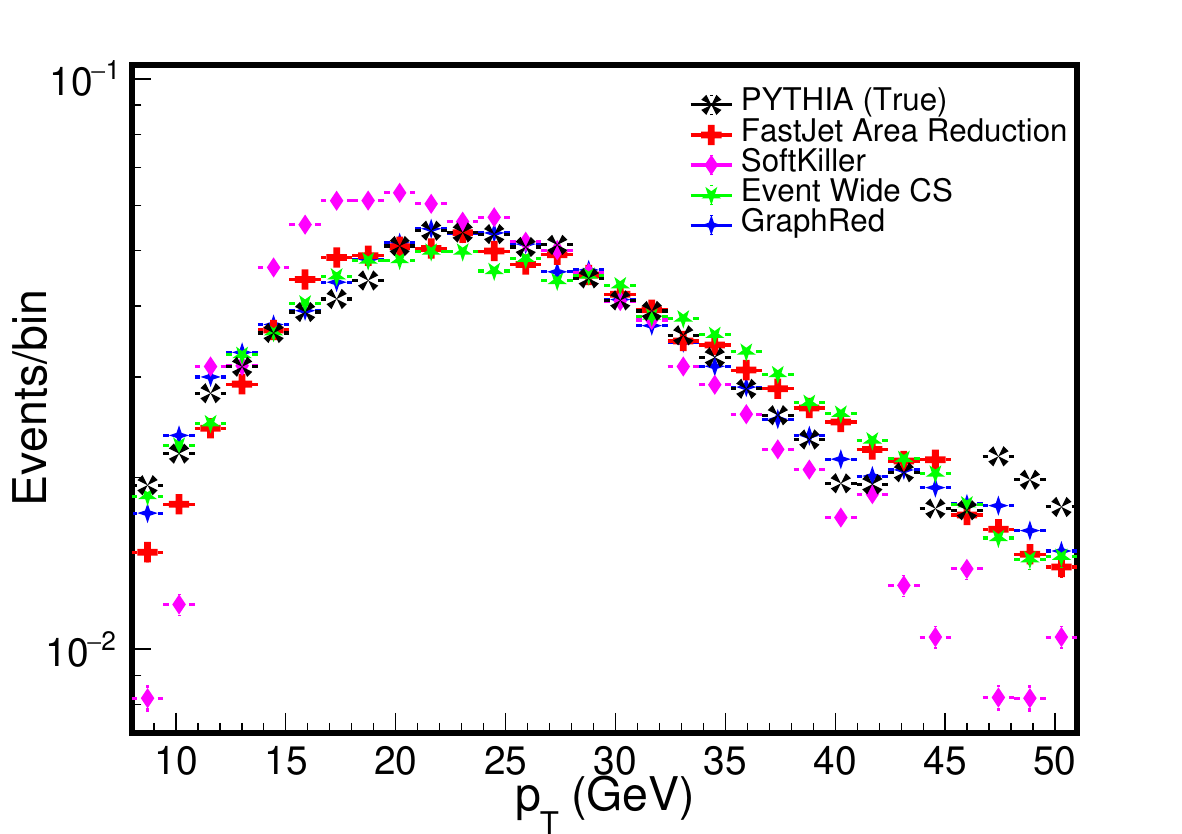}}
\end{minipage}%
\begin{minipage}{.5\linewidth}
\centering
\subfloat[$ E_{T}$ for leading jet]{\label{main:b}\includegraphics[scale=.4]{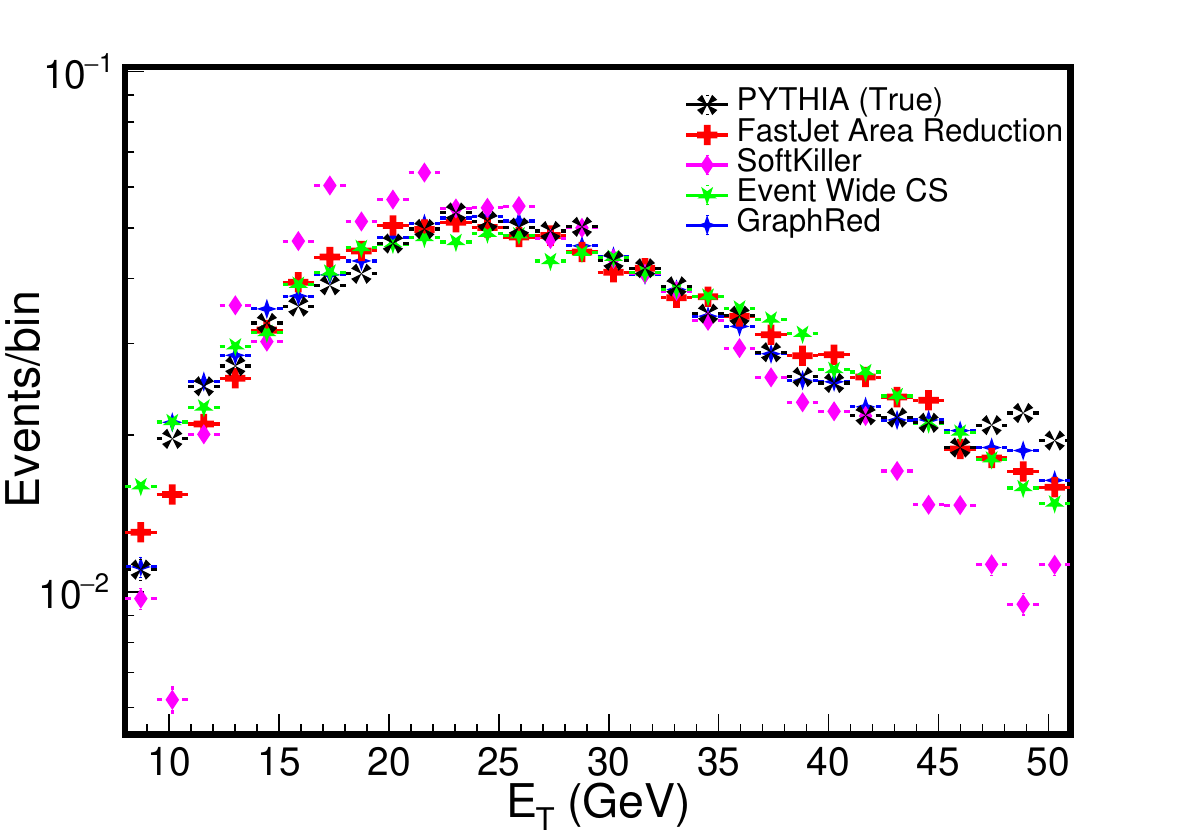}}
\end{minipage}\par\medskip

\caption{(color online) $p_{T}$ and $ E_{T}$ distribution for leading jet in embedded event dataset. The event compared with the background reduction techniques. The PYTHIA (True) is the true distribution obtained by clustering of hard particles. GraphRed distribution is the distribution obtained by clustering the hard particles predicted by the model.}
\label{fig:ljet_pythia}
\end{figure}

Figure~\ref{fig:jet_show} depicts the particles of an entire event in $(\eta, \phi)$ plane weighted with $p_T$, (a) all particles of jets, (b) and the particle of jet with leading $p_T$ (c). This is to demonstrate the various stage of jets, starting from events to final leading $p_T$ jets. To evaluate the performance of background reduction methods, we first compare the $p_{T}$ and $E_{T}$ for the leading jet in the event as reconstructed by all the methods discussed above. 
The $p_{T}$ and $E_{T}$ distributions are shown in Fig.~\ref{fig:ljet_nq}, \ref{fig:ljet_q} for non-quenched dataset and quenched dataset from full heavy-ion event from HYDJET whereas for embedded event by PYTHIA $\&$ HYDJET is shown in Fig.~\ref{fig:ljet_pythia}. 
We infer that GraphRed is able to reconstruct the jet with greater accuracy over other methods by preserving the shape of the distribution and modeling tail distribution also. It can also be seen from Fig.~\ref{fig:ljet_nq}, \ref{fig:ljet_q}, that GraphRed is able to model the low $p_{T}$ distribution shape whereas other methods fails to do this and deteriorates in this range. Moreover, at higher $p_{T}$ and  $E_{T}$ ranges GraphRed appears to differ from the true distribution, but the deviation is not significant as compared to other methods, which appears to be one limitation of this model and gives scope for future studies and optimization of model. Moreover, these comparisons also describe the physics objective behind the miss-characterization of hard particles. They will affect the shape of the distribution, which may affect the further interpretation and studies over it.\\

For an event by event comparison, we compare kinematic parameters like $E_{T}$, $p_{T}$ and spatial parameter like $\eta$ for the leading jet in an event. For each event, ratio of $E_{T}^{reco}/E_{T}^{true}$, $ p_{T}^{reco}/p_{T}^{true}$ and $ \Delta\eta$ ($\eta^{true}-\eta^{reco}$) is being computed for all the methods for the leading jet in an event.\\

\begin{figure}[!hbt]

\begin{minipage}{.5\linewidth}
\centering
\subfloat[$E_{T}^{reco}/E_{T}^{true}$ for non-quenched dataset of full heavy-ion event by HYDJET]{\label{main:a2}\includegraphics[scale=.35]{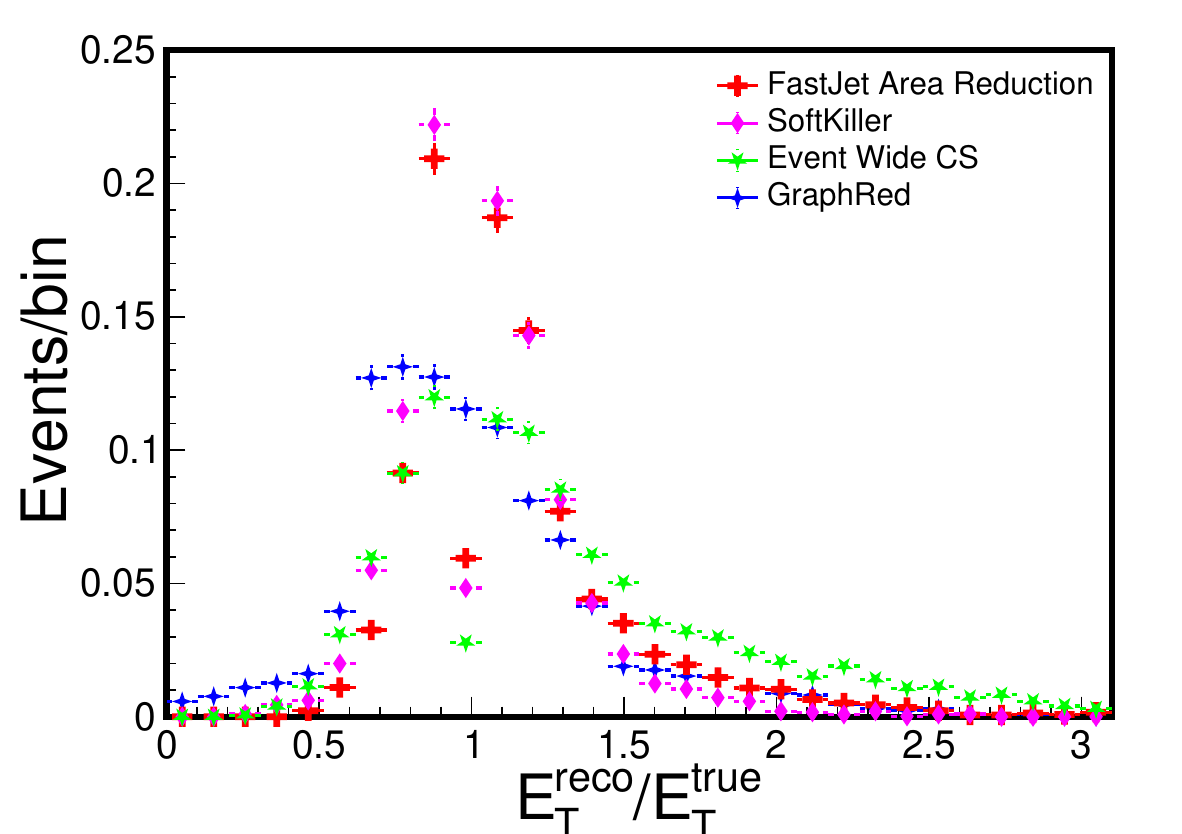}}
\end{minipage}%
\begin{minipage}{.5\linewidth}
\centering
\subfloat[$E_{T}^{reco}/E_{T}^{true}$ for quenched dataset of full heavy-ion event by HYDJET]{\label{main:a2}\includegraphics[scale=.35]{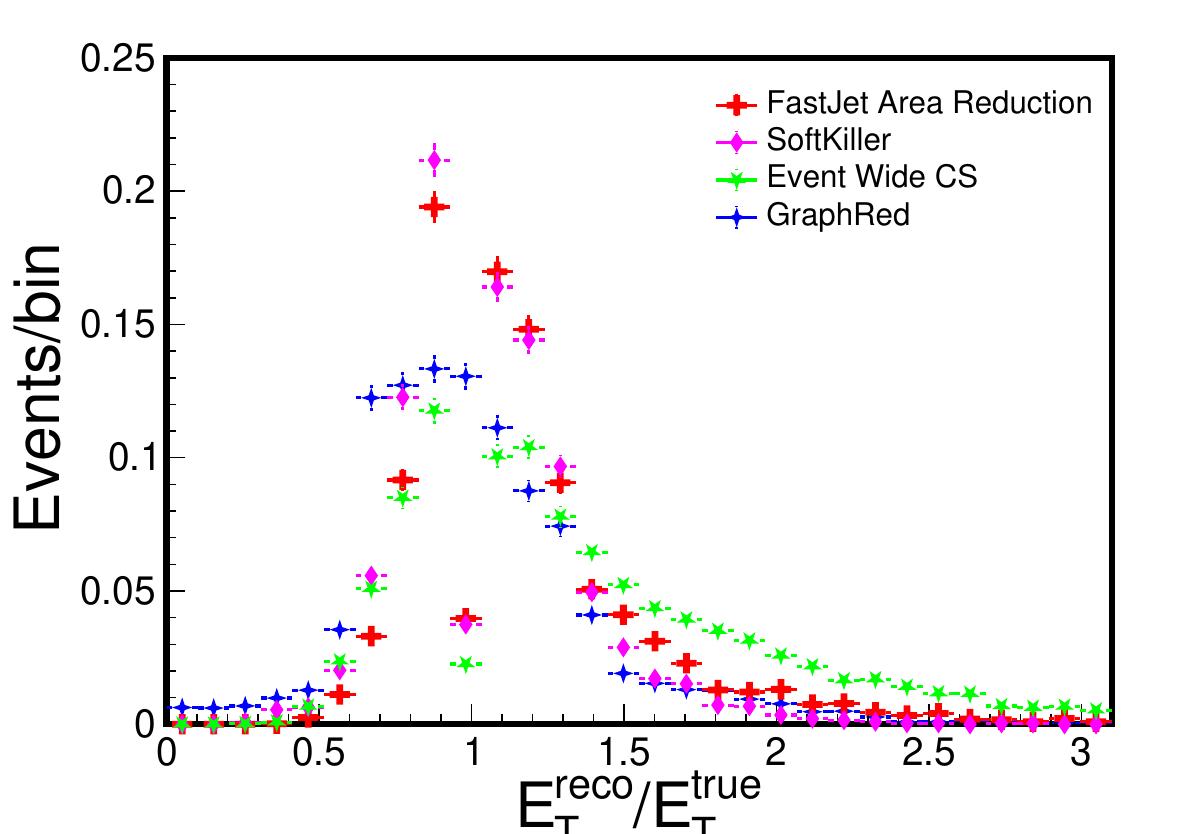}}
\end{minipage}
\begin{minipage}{.5\linewidth}
\centering
\hbox{\hspace{10.0em}\subfloat[$E_{T}^{reco}/E_{T}^{true}$ for embedded event dataset by PYTHIA $\&$ HYDJET]{\label{main:b2}\includegraphics[scale=.35]{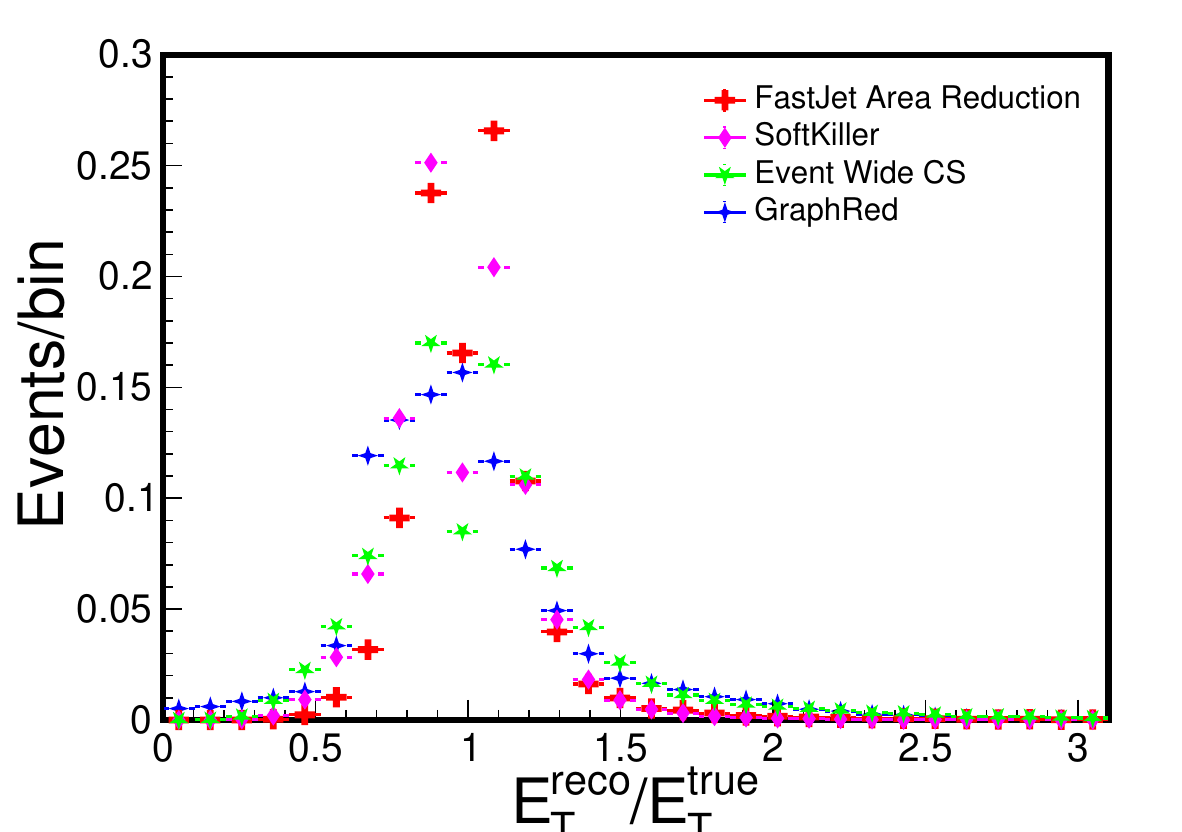}}}
\end{minipage}\par\medskip
\caption{(color online) $ E_{T}^{reco}/E_{T}^{true}$ comparison with various background subtraction methods }
\label{fig:et}
\end{figure}
In Fig.~\ref{fig:et}, $ E_{T}^{reco}/E_{T}^{true}$ is plotted for both data-sets. Ideally, a perfect method ratio must peak at one; however, the ratios are distributed around unity. It is clear from the ratio plots there are jagged peaks and dips present in the case of other methods, whereas for GraphRed range of peak is accumulated around unity representing a better quality of jet reconstruction for all the three plots representing a steady performance over them. We also observed that there are some events where GraphRed over- and under-predict the $E_{T}$, but these events are less frequent as compared with other methods.\\

\begin{figure}[!hbt]

\begin{minipage}{.5\linewidth}
\centering
\subfloat[$p_{T}^{reco}/p_{T}^{true}$ for non-quenched dataset of full heavy-ion event by HYDJET]{\label{main:a2}\includegraphics[scale=.35]{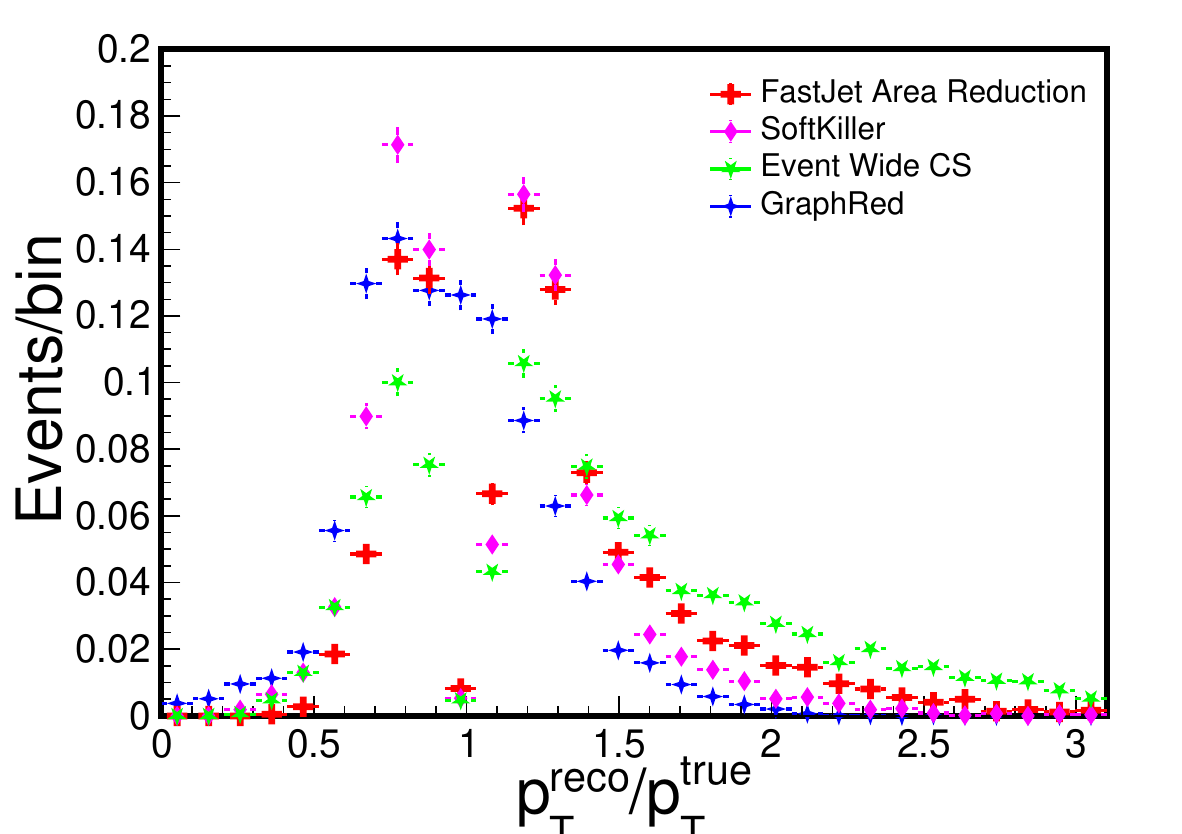}}
\end{minipage}%
\begin{minipage}{.5\linewidth}
\centering
\subfloat[$p_{T}^{reco}/p_{T}^{true}$ for quenched dataset of full heavy-ion event by HYDJET]{\label{main:a2}\includegraphics[scale=.35]{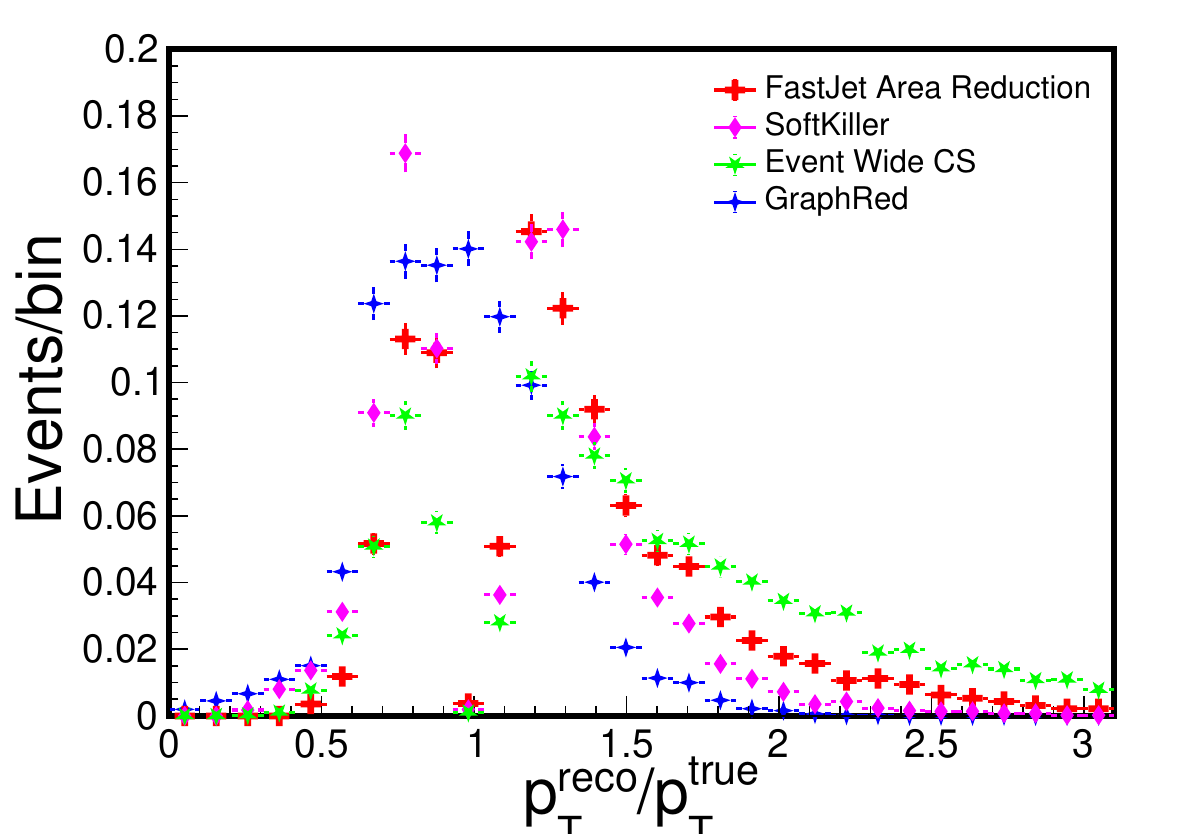}}
\end{minipage}
\begin{minipage}{.5\linewidth}
\centering
\hbox{\hspace{10.0em}\subfloat[$p_{T}^{reco}/p_{T}^{true}$ for embedded event dataset by PYTHIA $\&$ HYDJET]{\label{main:b2}\includegraphics[scale=.35]{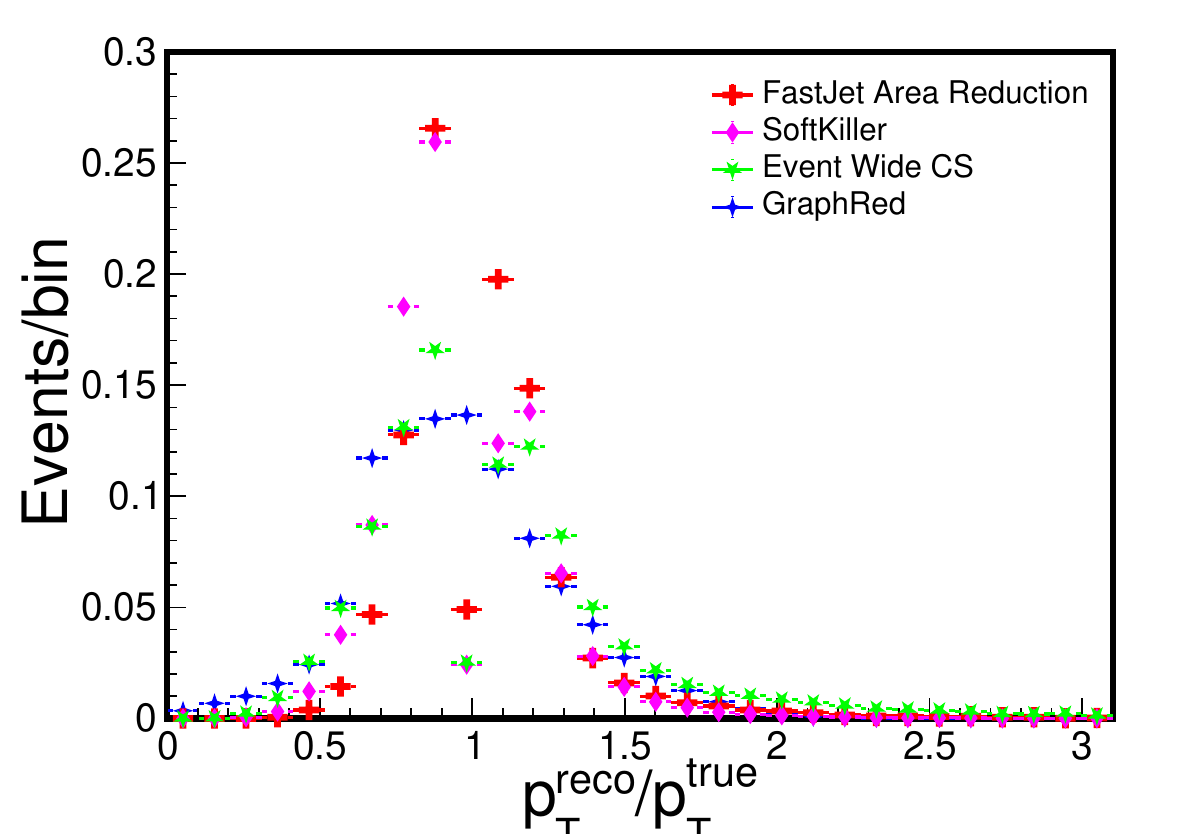}}}
\end{minipage}\par\medskip
\caption{(color online) $ p_{T}^{reco}/p_{T}^{true}$ comparison with various background subtraction methods }
\label{fig:pt}
\end{figure}

In Fig.~\ref{fig:pt}, $ p_{T}^{reco}/p_{T}^{true}$ is plotted for both data-sets. We observe a similar behavior of the methods as compared to $ E_{T}^{reco}/E_{T}^{true}$ ratio plot. The usage of GraphRed as a pre-processing method led to the ratio peaks around unity as compared to other methods for whom we observe peaks and dips around the unit. These methods suffer from a little over-or under-prediction of $p_{T}$ and $E_{T}$, which can be seen from the tail distribution of various methods. 
\newpage
\begin{figure}[!hbt]

\begin{minipage}{.5\linewidth}
\centering
\subfloat[$\Delta\eta$ for non-quenched dataset of full heavy-ion event by HYDJET]{\label{main:a2}\includegraphics[scale=.35]{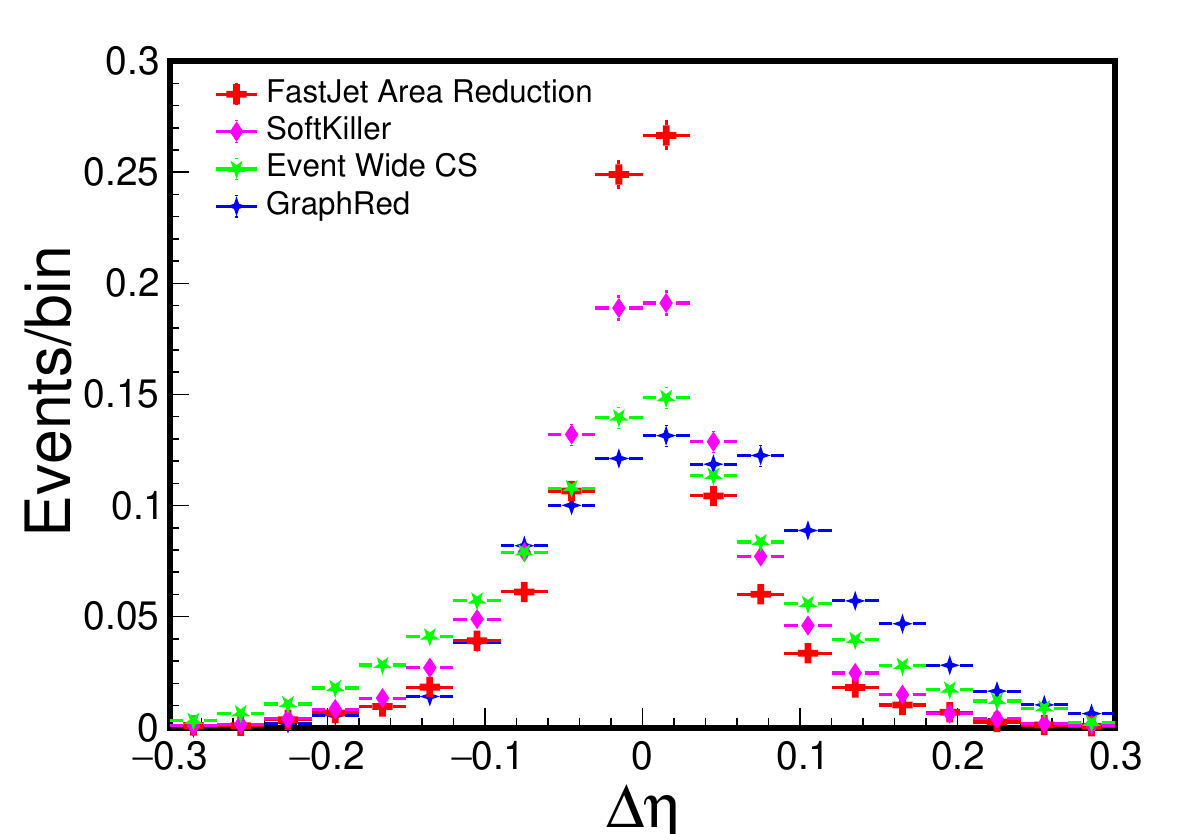}}
\end{minipage}%
\begin{minipage}{.5\linewidth}
\centering
\subfloat[$\Delta\eta$ for quenched dataset of full heavy-ion event by HYDJET]{\label{main:a2}\includegraphics[scale=.35]{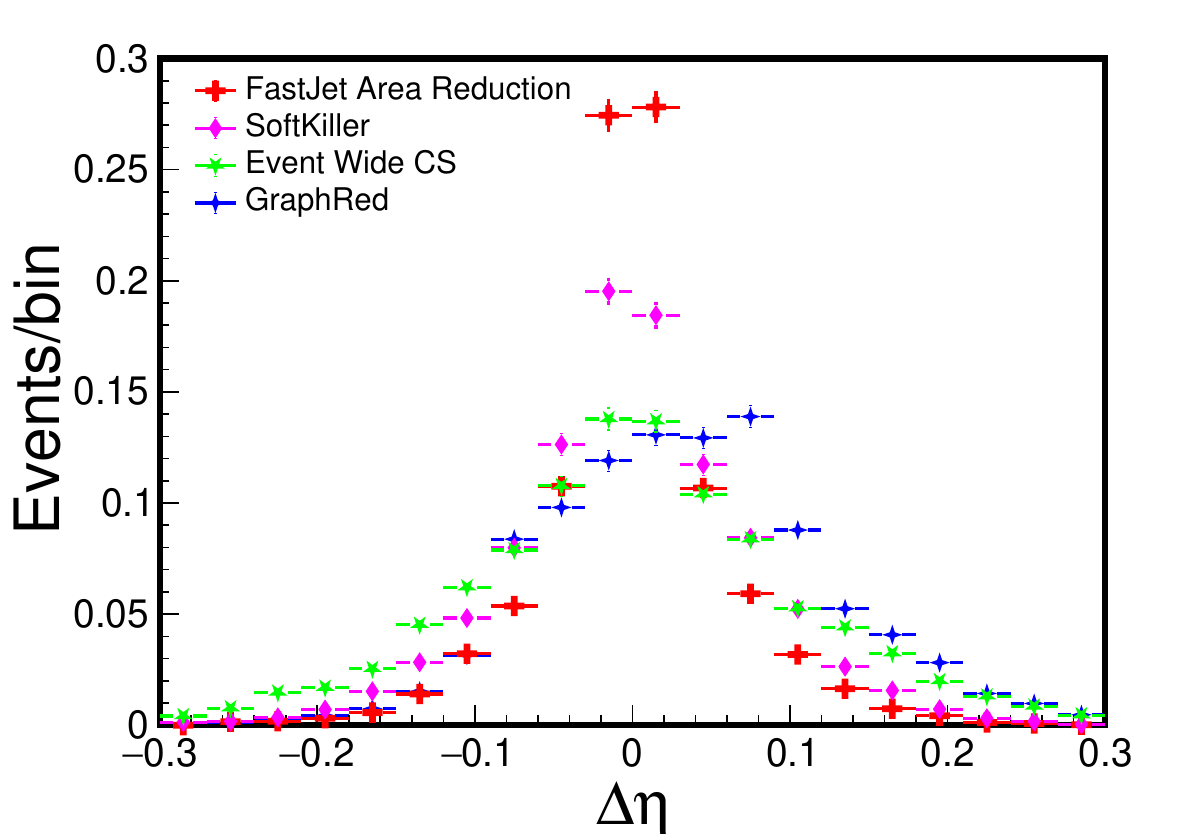}}
\end{minipage}
\begin{minipage}{.5\linewidth}
\centering
\hbox{\hspace{10.0em}\subfloat[$\Delta\eta$ for embedded event dataset by PYTHIA $\&$ HYDJET]{\label{main:b2}\includegraphics[scale=.35]{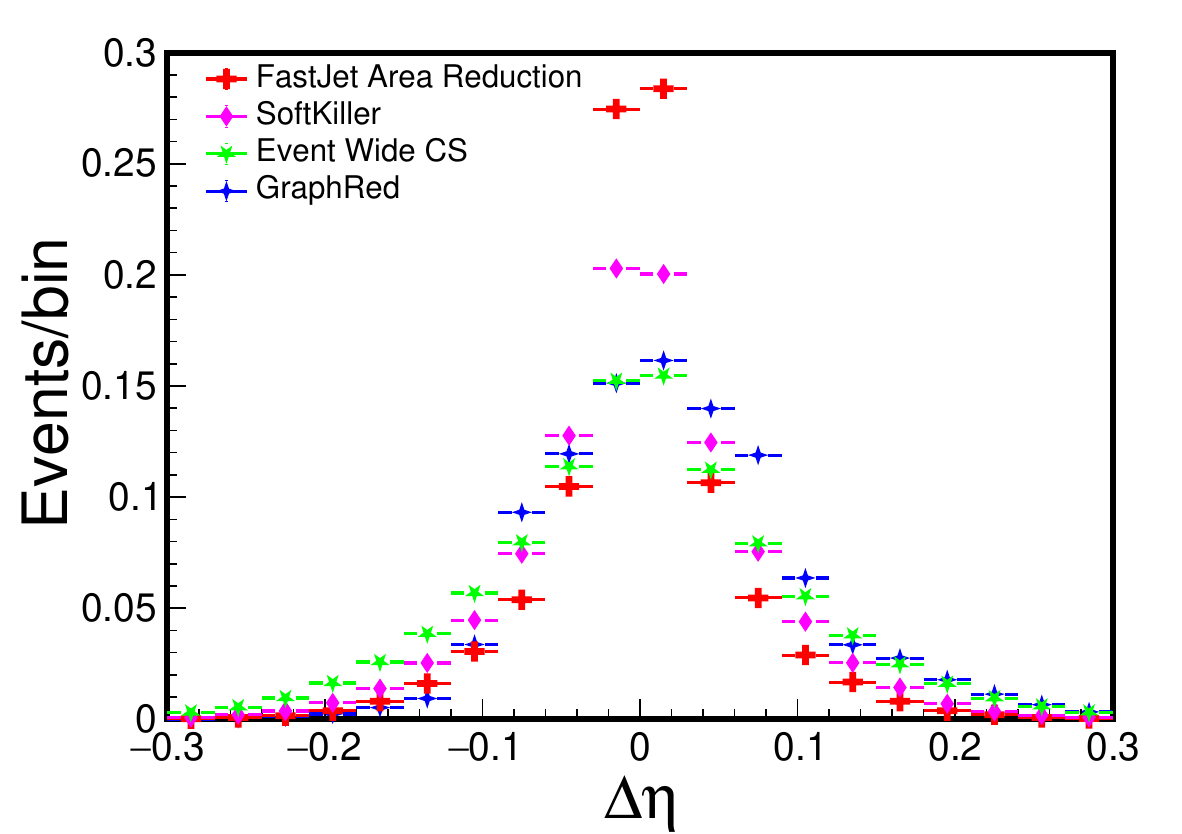}}}
\end{minipage}\par\medskip
\caption{(color online) $\Delta\eta$ comparison with various background subtraction methods }
\label{fig:eta}
\end{figure}

In Fig.~\ref{fig:eta}, $ \Delta\eta$ ($\eta^{reco}-\eta^{true}$) distribution is plotted. We observe that FastJet Area Reduction and SoftKiller perform better than GraphRed in the spatial resolution, which appears to be a limitation of the model, but it is worthy to note that these methods did not perform equivalently well in the case of $p_{T}$ and $E_{T}$. It also important to note that GraphRed is a pre-processing method that only identifies the most-likely hard particles. 

\begin{figure}[!hbt]

\begin{minipage}{.5\linewidth}
\centering
\subfloat[$<\Delta p_{T}>$ for non-quenched dataset of full heavy-ion event by HYDJET]{\label{main:a2}\includegraphics[scale=.35]{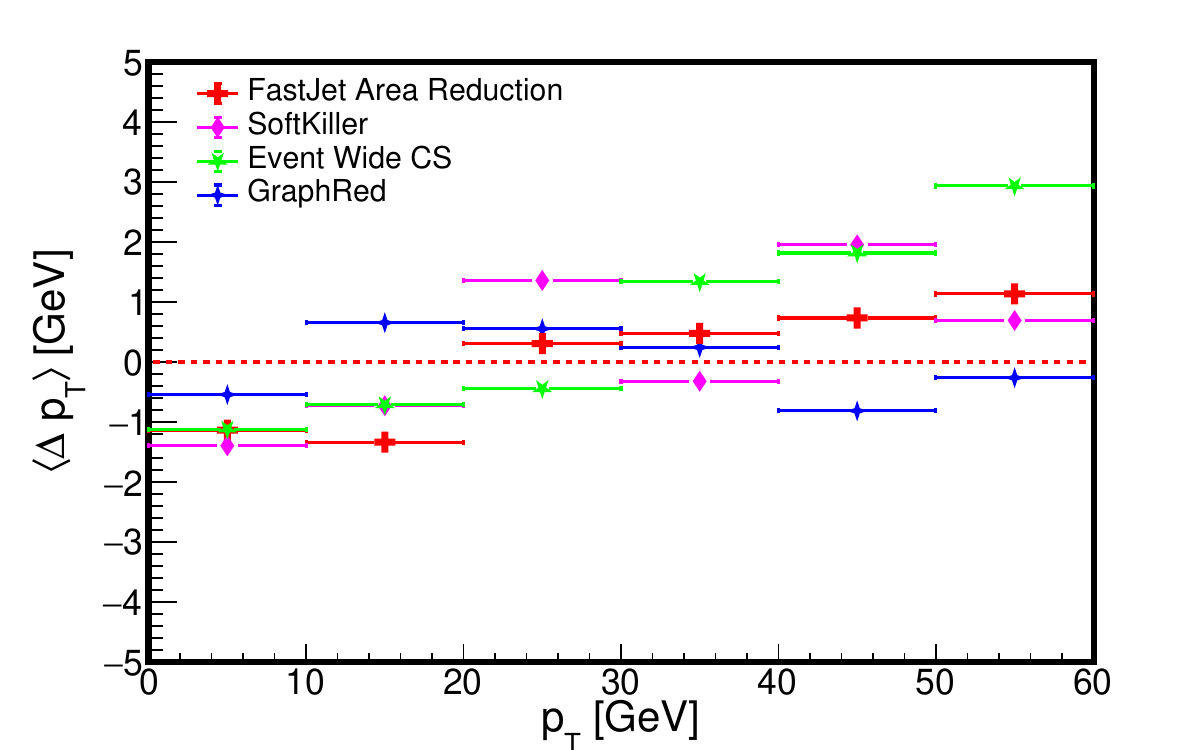}}
\end{minipage}%
\begin{minipage}{.5\linewidth}
\centering
\subfloat[$<\Delta p_{T}>$ for quenched dataset of full heavy-ion event by HYDJET]{\label{main:a2}\includegraphics[scale=.35]{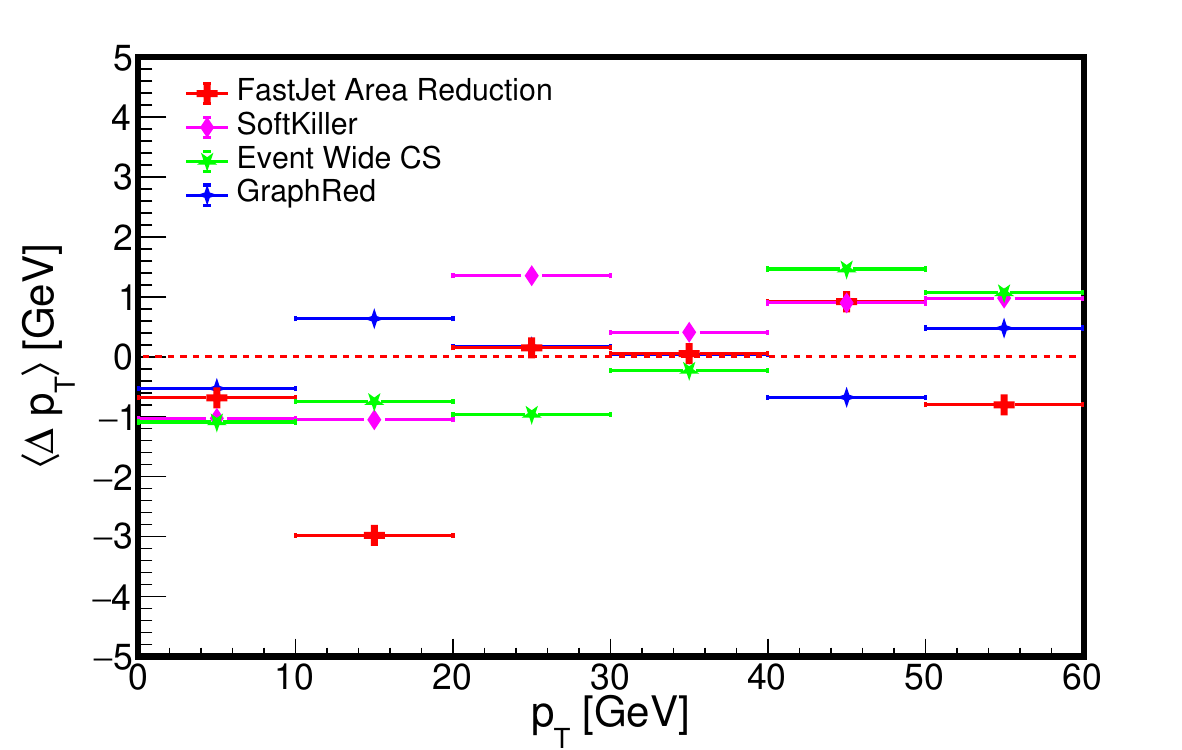}}
\end{minipage}
\begin{minipage}{.5\linewidth}
\centering
\hbox{\hspace{10.0em}\subfloat[$<\Delta p_{T}>$ for embedded event dataset by PYTHIA $\&$ HYDJET]{\label{main:b2}\includegraphics[scale=.35]{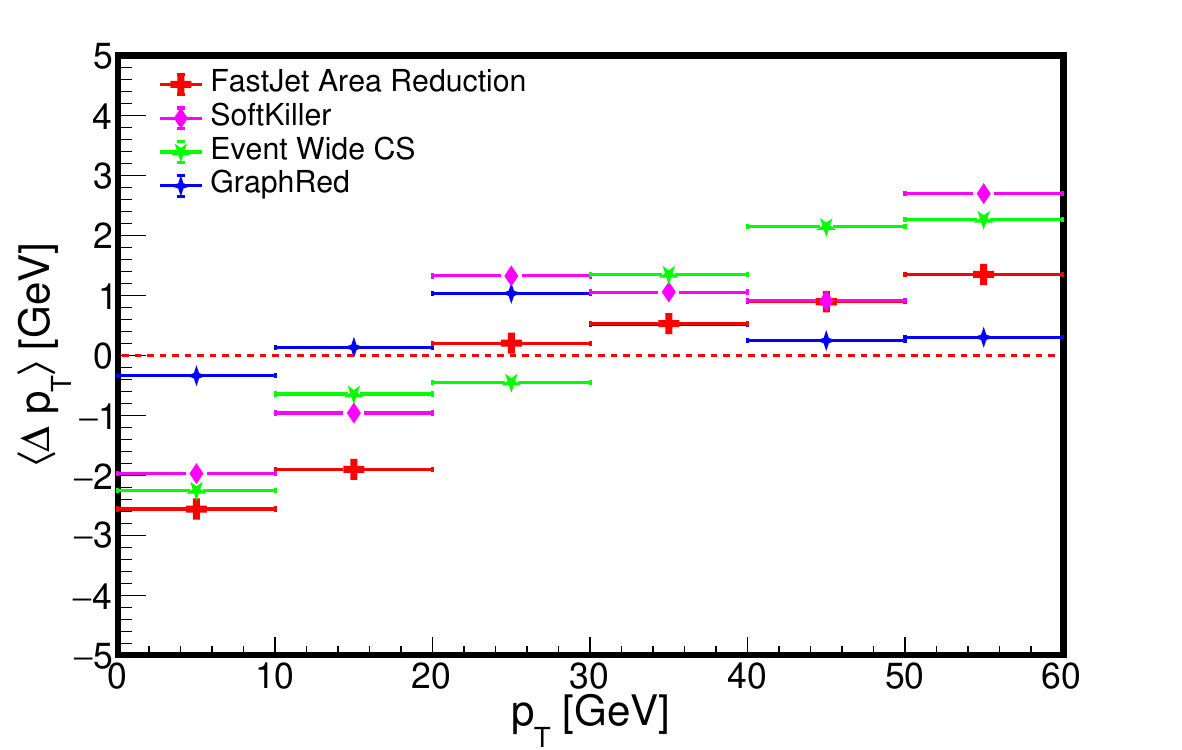}}}
\end{minipage}\par\medskip
\caption{(color online) $<\Delta p_{T}>$ comparison with various background subtraction methods }
\label{fig:average_pt}
\end{figure}

\begin{figure}[!hbt]

\begin{minipage}{.5\linewidth}
\centering
\subfloat[$<\Delta E_{T}>$ for non-quenched dataset of full heavy-ion event by HYDJET]{\label{main:a2}\includegraphics[scale=.35]{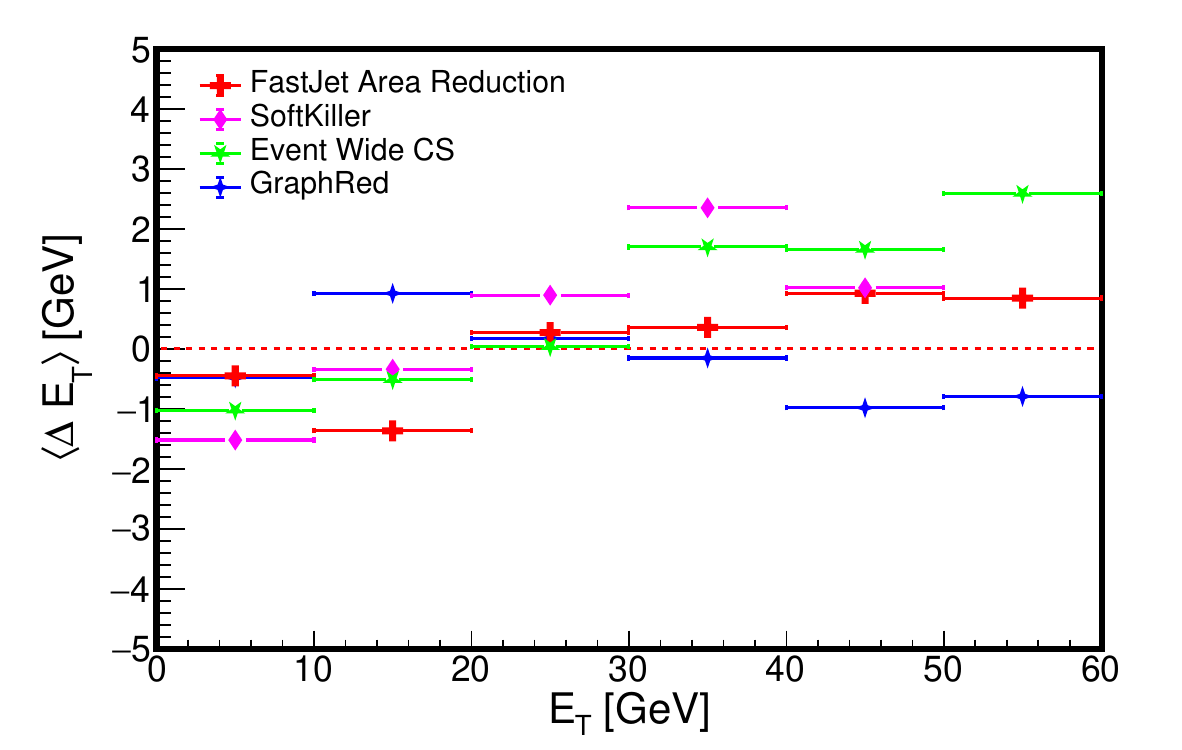}}
\end{minipage}%
\begin{minipage}{.5\linewidth}
\centering
\subfloat[$<\Delta E_{T}>$ for quenched dataset of full heavy-ion event by HYDJET]{\label{main:a2}\includegraphics[scale=.35]{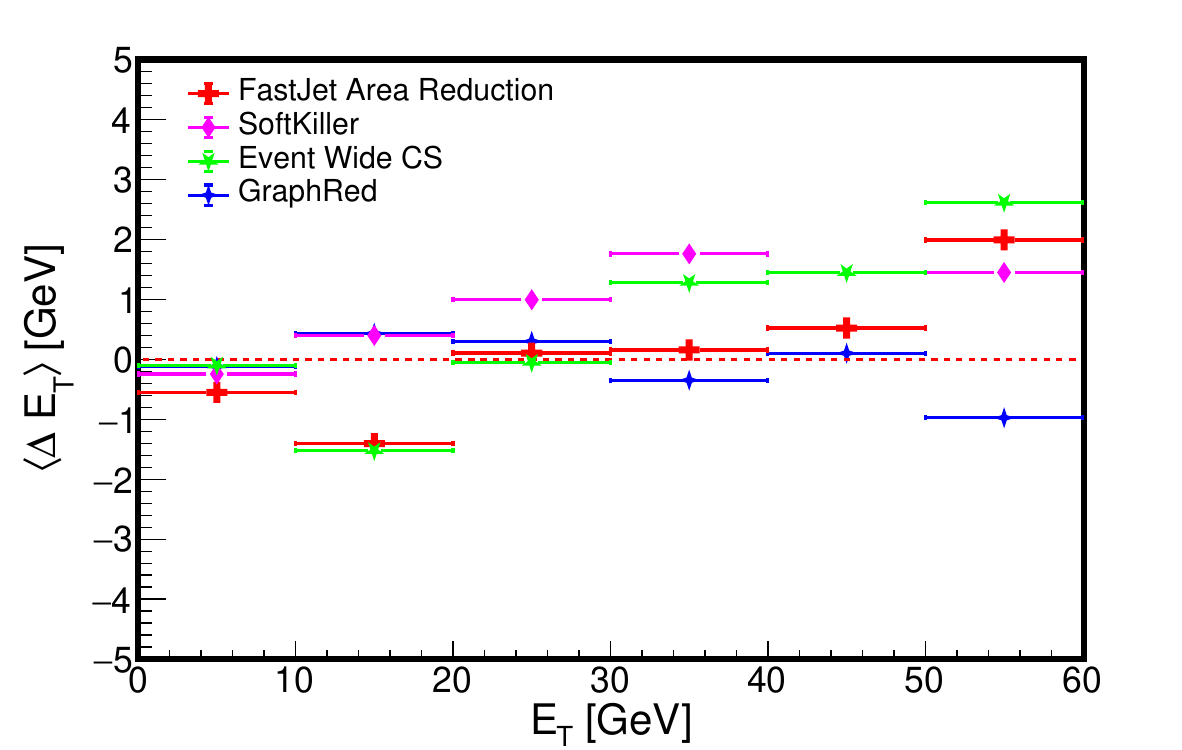}}
\end{minipage}
\begin{minipage}{.5\linewidth}
\centering
\hbox{\hspace{10.0em}\subfloat[$<\Delta E_{T}>$ for embedded event dataset by PYTHIA $\&$ HYDJET]{\label{main:b2}\includegraphics[scale=.35]{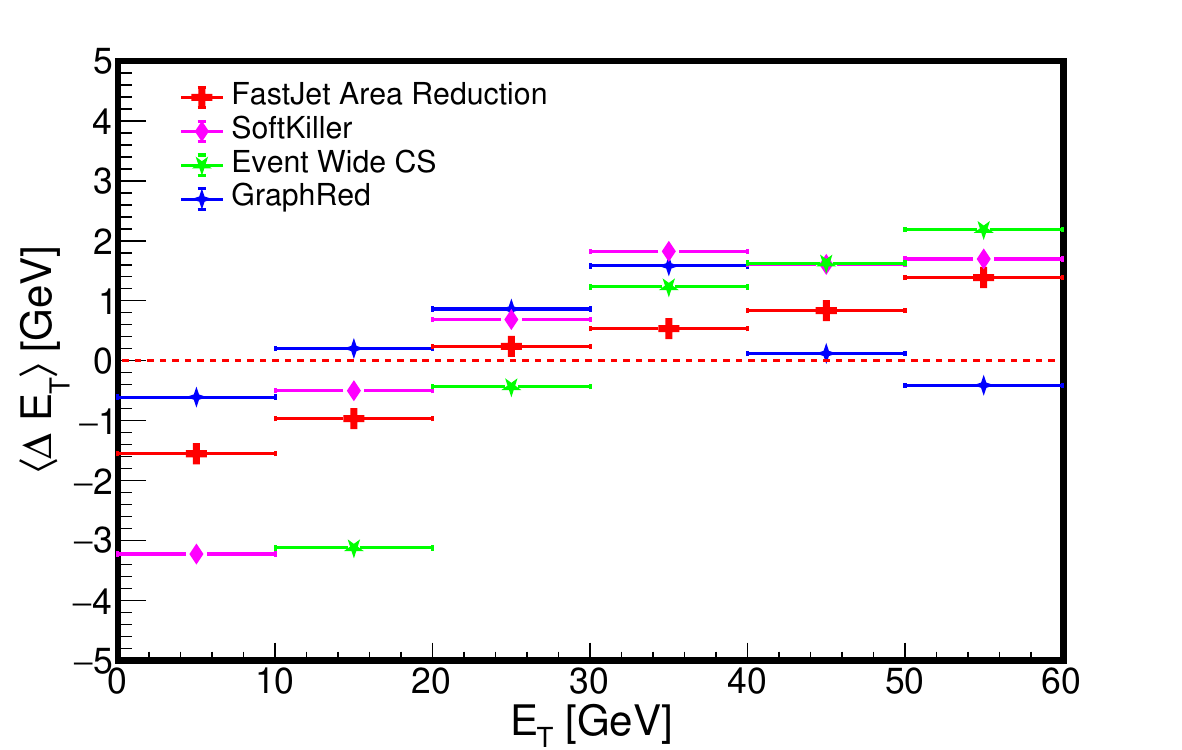}}}
\end{minipage}\par\medskip
\caption{(color online) $<\Delta E_{T}>$ comparison with various background subtraction methods }
\label{fig:average_et}
\end{figure}

\begin{figure}[!hbt]

\begin{minipage}{.5\linewidth}
\centering
\subfloat[$<\Delta \eta>$ for non-quenched dataset of full heavy-ion event by HYDJET]{\label{main:a2}\includegraphics[scale=.35]{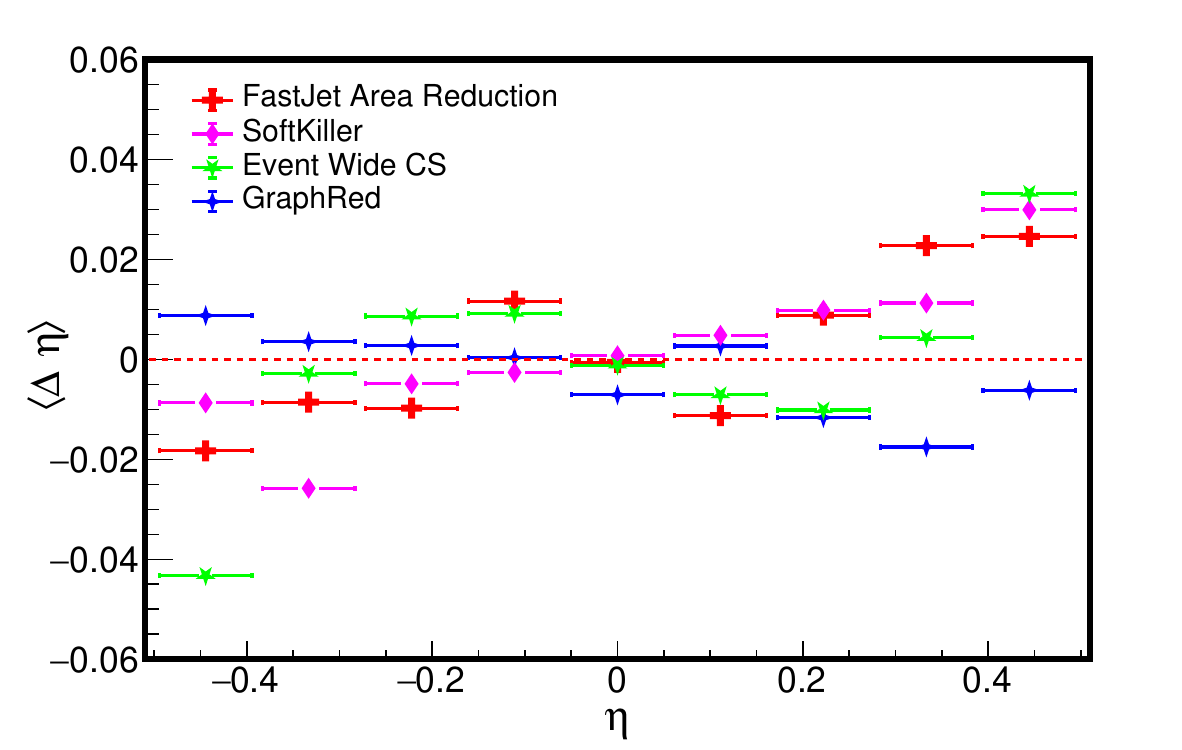}}
\end{minipage}%
\begin{minipage}{.5\linewidth}
\centering
\subfloat[$<\Delta \eta>$ for quenched dataset of full heavy-ion event by HYDJET]{\label{main:a2}\includegraphics[scale=.35]{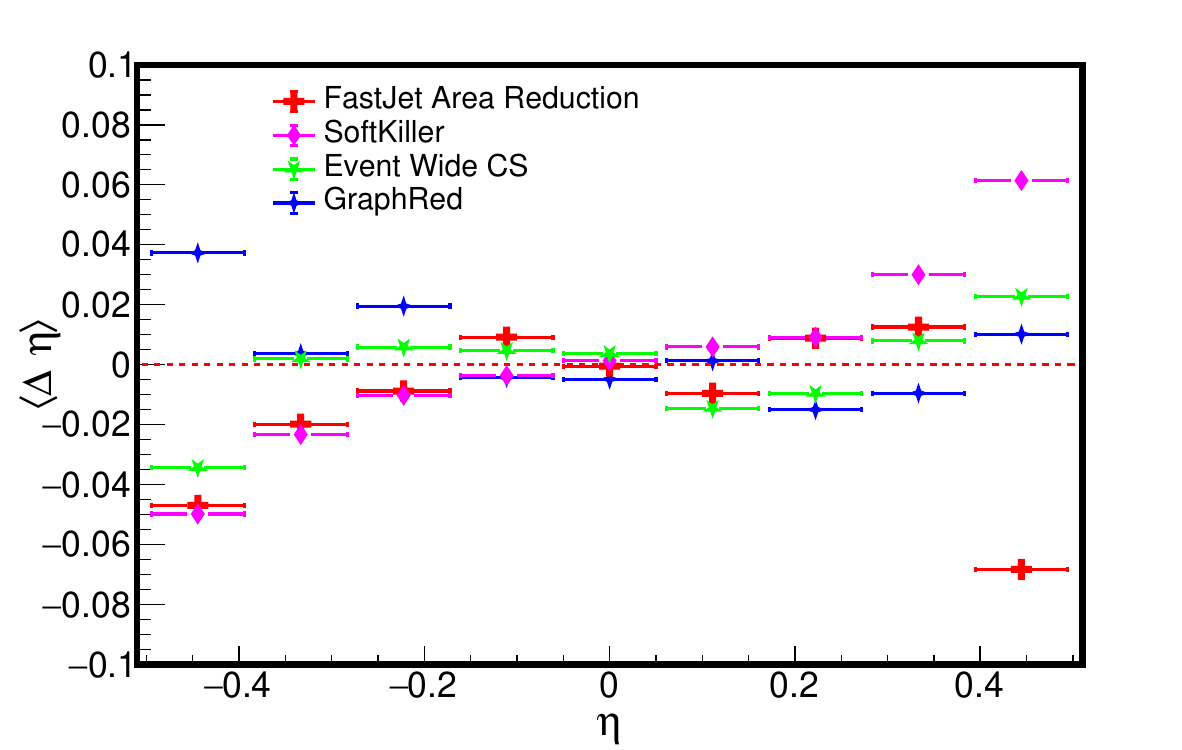}}
\end{minipage}
\begin{minipage}{.5\linewidth}
\centering
\hbox{\hspace{10.0em}\subfloat[$<\Delta \eta>$ for embedded event dataset by PYTHIA $\&$ HYDJET]{\label{main:b2}\includegraphics[scale=.35]{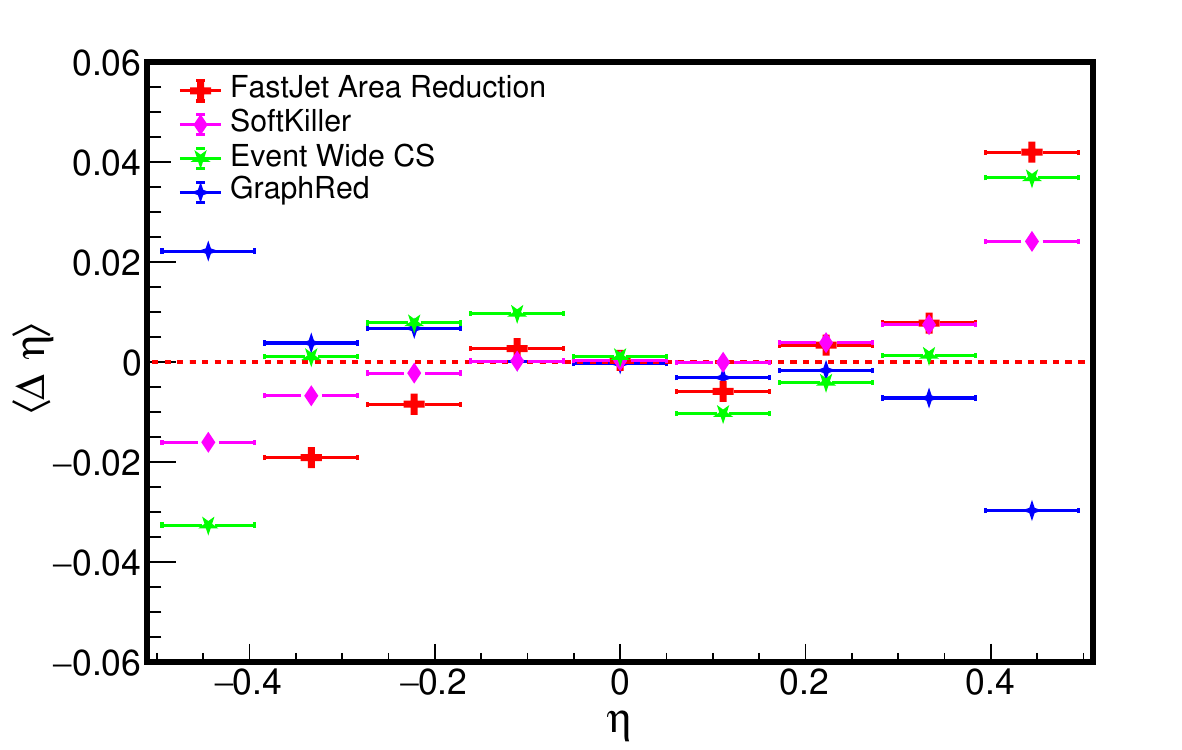}}}
\end{minipage}\par\medskip
\caption{(color online) $<\Delta \eta>$ comparison with various background subtraction methods }
\label{fig:average_eta}
\end{figure}

We also compared the average shift ($<\Delta p_{T}>$, $<\Delta E_{T}>$ and $<\Delta \eta>$) for the leading jet as a function of respective ranges among different methods where
\begin{subequations}
\begin{align}
  \Delta p_{T} &= p_{T}^{true} - p_{T}^{reco}\\
  \Delta E_{T} &= E_{T}^{true} - E_{T}^{reco}\\
  \Delta \eta &= \eta^{true} - \eta^{reco}
\end{align}
\end{subequations}
The average shift in $p_{T}$ i.e. $<\Delta p_{T}>$ as a function of $p_{T}$ range is shown in Fig.~\ref{fig:average_pt},$<\Delta E_{T}>$ as a function of $E_{T}$ range is shown in Fig.~\ref{fig:average_et} and $<\Delta \eta>$ as a function of $\eta$ range is shown in Fig.~\ref{fig:average_eta} for both full heavy-ion events by HYDJET and embedded events by PYTHIA $\&$ HYDJET. 
\newpage
The plots for average shift in $p_{T}$ ($<\Delta p_{T}>$) is shown in Fig.~\ref{fig:average_pt} for both datasets. It can be seen that for GraphRed, the average shift mostly remains in the range $\in [-1,1]$ GeV. On the contrary, the shifts have significant offsets when computed from the other methods and deviate beyond the $[-1,1]$ GeV range, not performing better than GraphRed. Sometimes, other methods like FastJet Area Reduction, SoftKiller performs better than GraphRed in a local regime, but GraphRed dominates over the other methods in the global regime. This is because the addition of background can significantly affect the clustering of particles as a proportion of constituents particles of a jet can be gained by or lost from the jet when
clustering the event with the full background.

The plots for average shift in $E_{T}$ ($<\Delta E_{T}>$) is shown in Fig.~\ref{fig:average_et} for both datasets. A similar behavior of the methods as compared to $<\Delta p_{T}>$ shift is observed here. The application of GraphRed as a pre-processing method to characterize hard and soft particles has led to a better resolution in $p_{T}$ and $E_{T}$ of the jets as compared to other methods. The other methods suffer from a large background present in the heavy-ion event and sometimes incorporate background contamination while clustering particles to make jets.\\
\newpage
The average shift distribution in $\eta$ ($<\Delta \eta>$) is shown in Fig.~\ref{fig:average_eta} for both datasets. We can observe that for $|\eta| > 0.4$, there appears a significant deviation in GraphRed and other methods. For $|\eta| < 0.4$, GraphRed performs equivalently well to other methods while limiting the offsets $\in [-0.02,0.02]$. The above plot also describes that a good spatial resolution can also be obtained using GraphRed and the energy and momentum resolution described previously.

\newpage
\section{Conclusion}
\label{sec:conc}

This article introduced GraphRed, a novel machine learning technique to find the jet constituent particles in heavy-ion collisions. The method is based on physics aware attention graph neural networks, which manifests a geometric deep learning technique to characterize jet in high multiplicity environments by appropriately separating the background. This method separates particles produced by hard and soft processes in the heavy-ion collisions system. By contrast, our model can work directly on particle-level observables in an event. The graph models combine the power of physics with attention graph neural networks with an efficient geometrical embedding of the particle spectra. We showed that this method performs equivalently well and substantially improves existing frameworks.  Our studies have revealed that geometrical embedding and creating a web graph for each particle during the construction played an essential role in the model's performance and we have also shown results in a gain in terms of accuracy and background suppression as compared to the traditional method. Moreover, for the first time, we show a model working on particle-level exploiting the latent representation with incorporating domain and topological knowledge from physics to segregate particles.

Being able to properly separate the products of the hard collisions from the softer background particles on an event-by-event basis, these results offer a solid direction for implementing essential machine learning-based algorithms in jet finding. It also shows its robustness to the event-by-event fluctuations, model-dependent effects present in the data, which is a notable application of machine learning on the data at the LHC energies. The work presented in the article opens the way to better comparisons between experimental data and theoretical calculation and provides a nascent step towards a new paradigm of efficient, robust, and tractable tools for jet physics in heavy-ion collisions at LHC. In conclusion, we believe that there exists a great potential in advanced techniques which incorporate physics knowledge and exploit the power of artificial intelligence.

\acknowledgments

The simulation and training works were carried-out in the computing facility in EHEP Lab at IISER Mohali. 

\bibliographystyle{unsrt}
\bibliography{Jet.bib}
\newpage
\appendix
\section{Training and Output distribution plots }
\subsection{Full heavy-ion event by HYDJET}

The training and validation accuracy w.r.t epochs are shown here for both quenched and non-quenched data-sets. We infer that an accuracy of 98-99$\%$ is achieved for both the cases, depicting a good performance of our GraphRed model.
\begin{figure}[!hbt]
\def\tabularxcolumn#1{m{#1}}
\begin{tabularx}{\linewidth}{@{}cXX@{}}
\begin{tabular}{cc}
\subfloat[Training Accuracy]{\hspace{-1.0em}\includegraphics[scale=0.41]{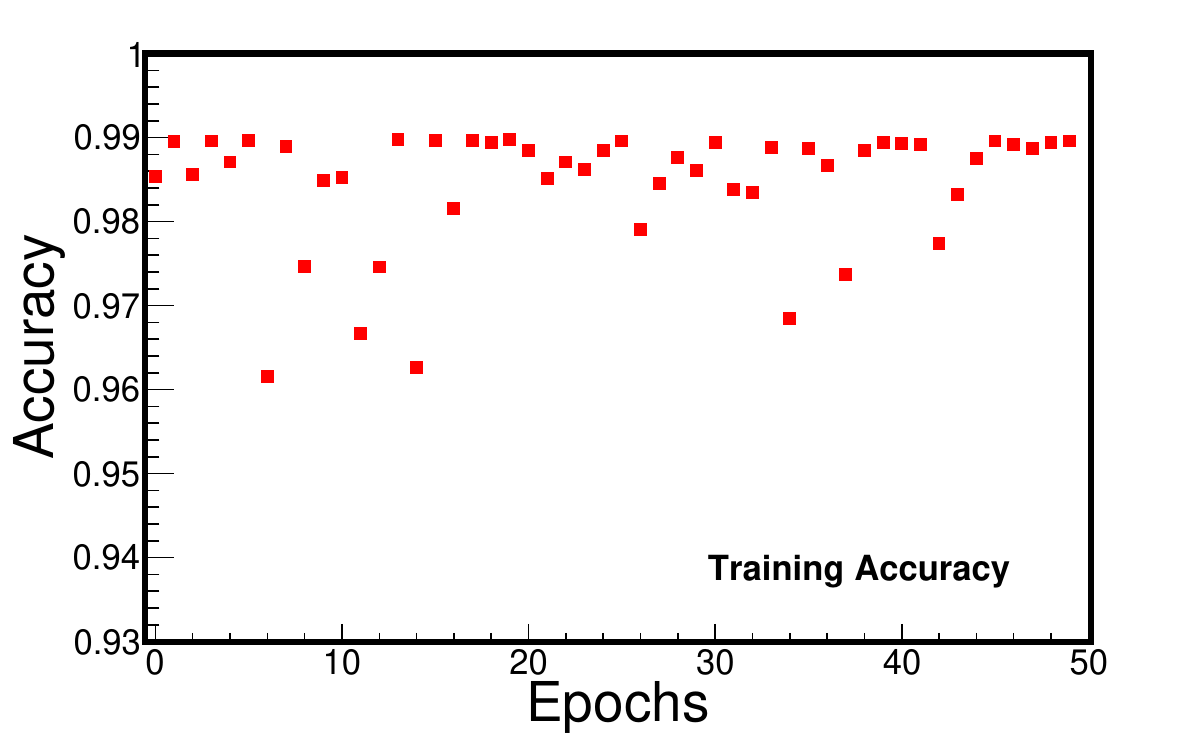}} 
   & \subfloat[Validation Accuracy]{\hspace{-2.0em}\includegraphics[scale=0.41]{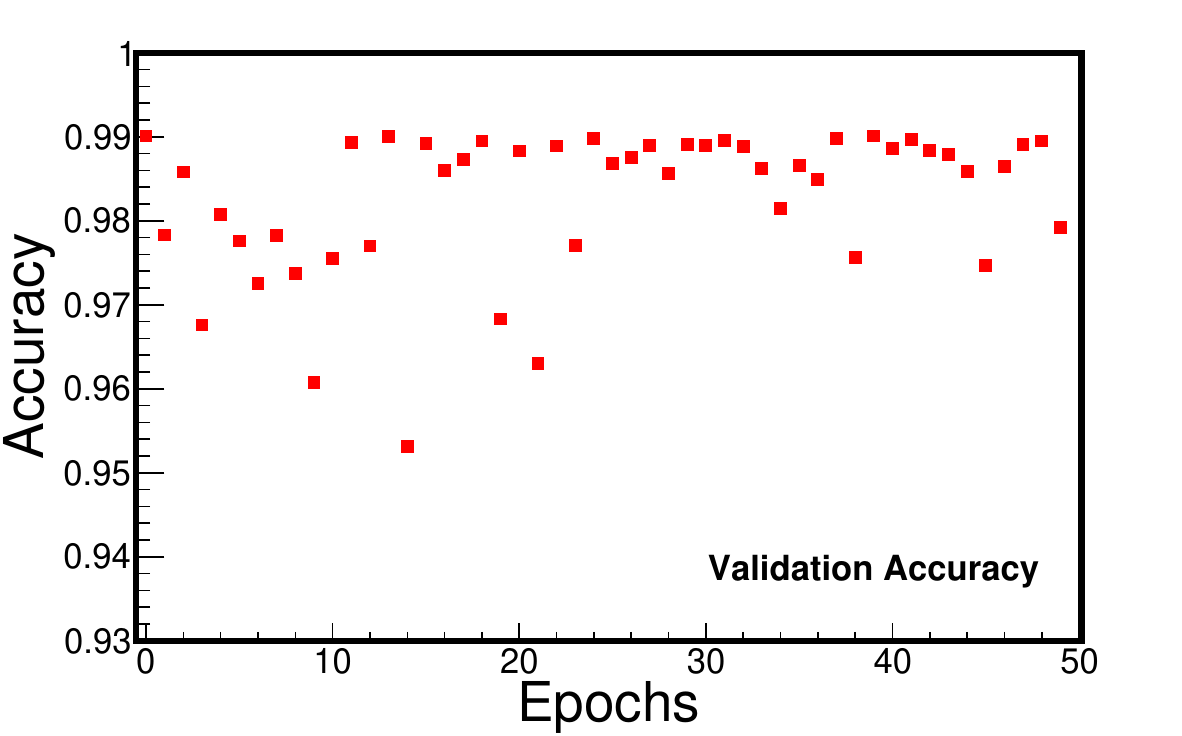}}\\
\end{tabular}
\end{tabularx}
\caption{(color online) Accuracy vs Epochs for Graph Neural Network for Non-Quenched Dataset}\label{fig:acc}
\end{figure}

\begin{figure}[!hbt]
\def\tabularxcolumn#1{m{#1}}
\begin{tabularx}{\linewidth}{@{}cXX@{}}
\begin{tabular}{cc}
\subfloat[Training Accuracy]{\hspace{-1.0em}\includegraphics[scale=0.41]{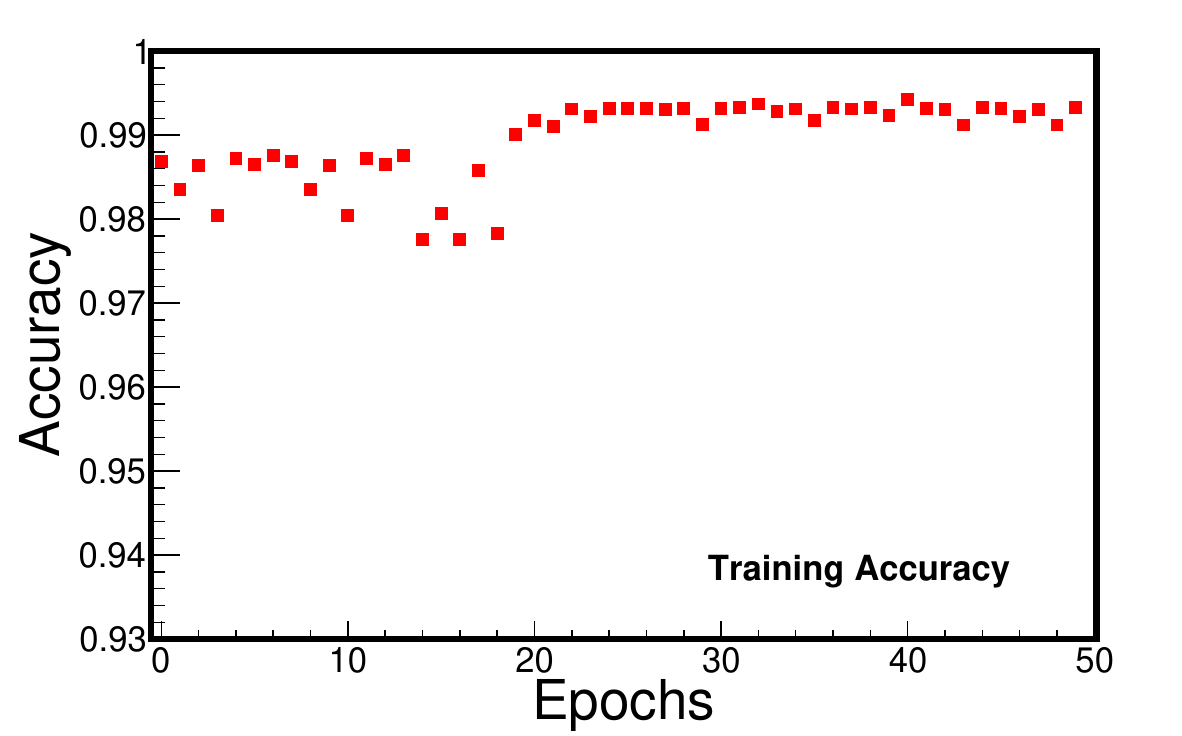}} 
   & \subfloat[Validation Accuracy]{\hspace{-2.0em}\includegraphics[scale=0.41]{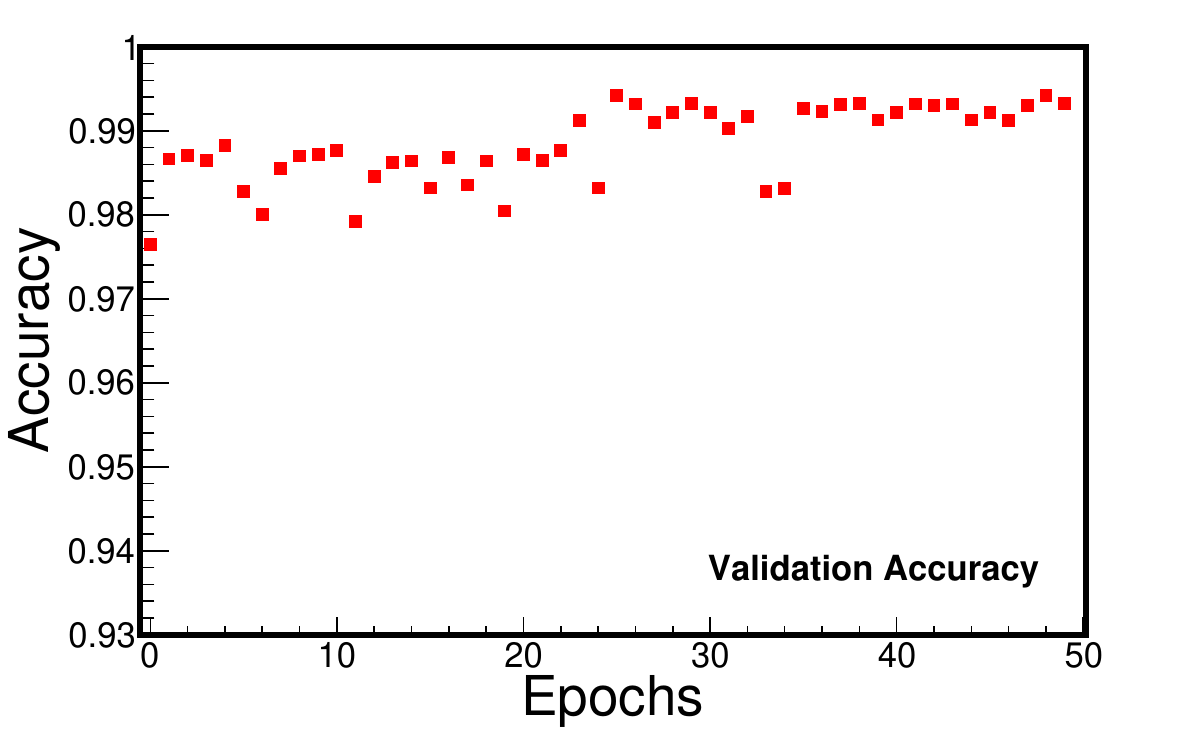}}\\
\end{tabular}
\end{tabularx}
\caption{(color online) Accuracy vs Epochs for Graph Neural Network for Quenched Dataset}\label{fig:acc2}
\end{figure}

The output probability distribution for GraphRed model is shown below for hard and soft particles for both quenched and non-quenched datasets. We observe that GraphRed is able to identify soft particles giving a peak around 0 and for hard particles there is a peak around 1 describing the correct identification alongside with miss-classification of some particles. 

\begin{figure}[!hbt]
\def\tabularxcolumn#1{m{#1}}
\begin{tabularx}{\linewidth}{@{}cXX@{}}
\begin{tabular}{cc}
\subfloat[Non-quenched dataset]{\hspace{-1.0em}\includegraphics[scale=0.41]{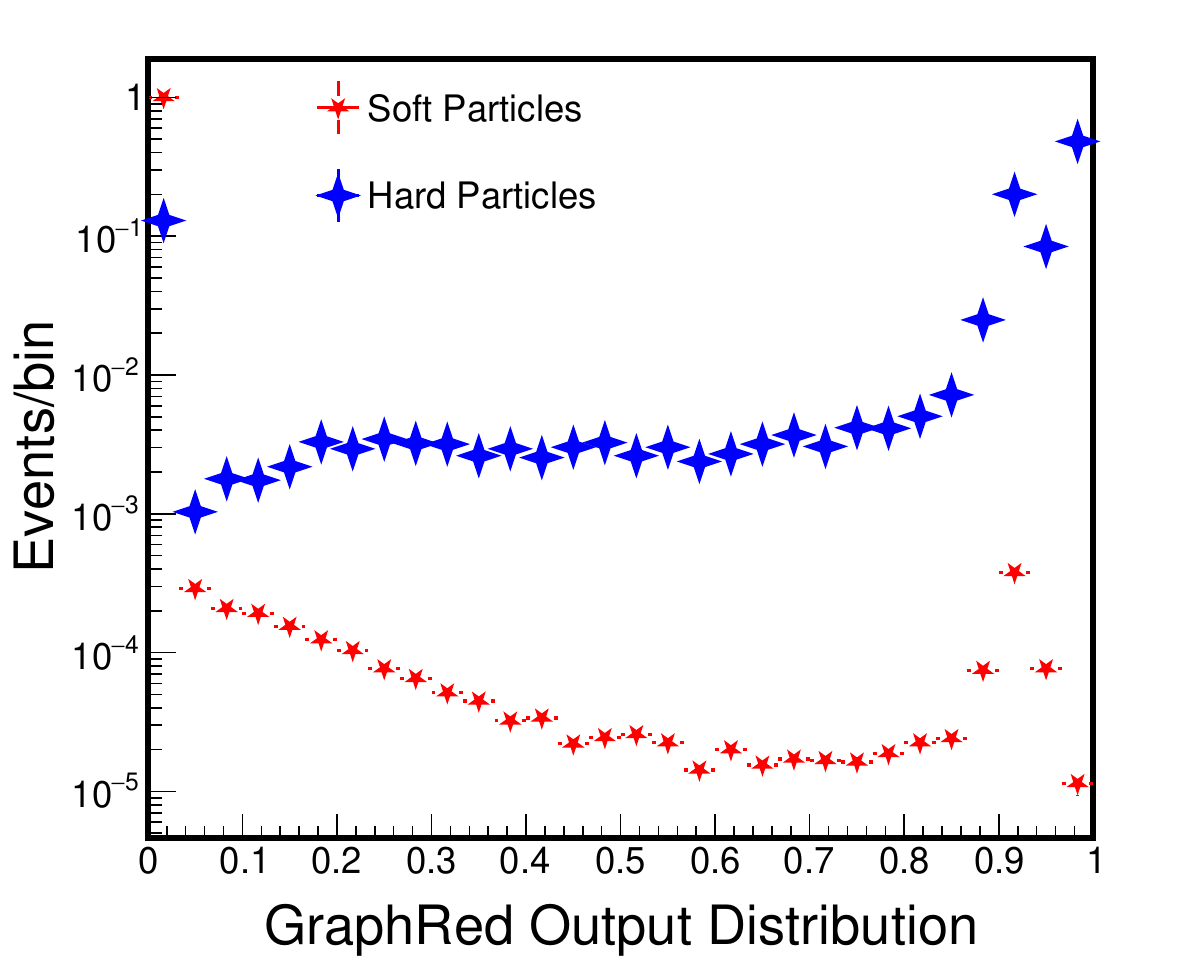}} 
   & \subfloat[quenched dataset]{\hspace{-2.0em}\includegraphics[scale=0.41]{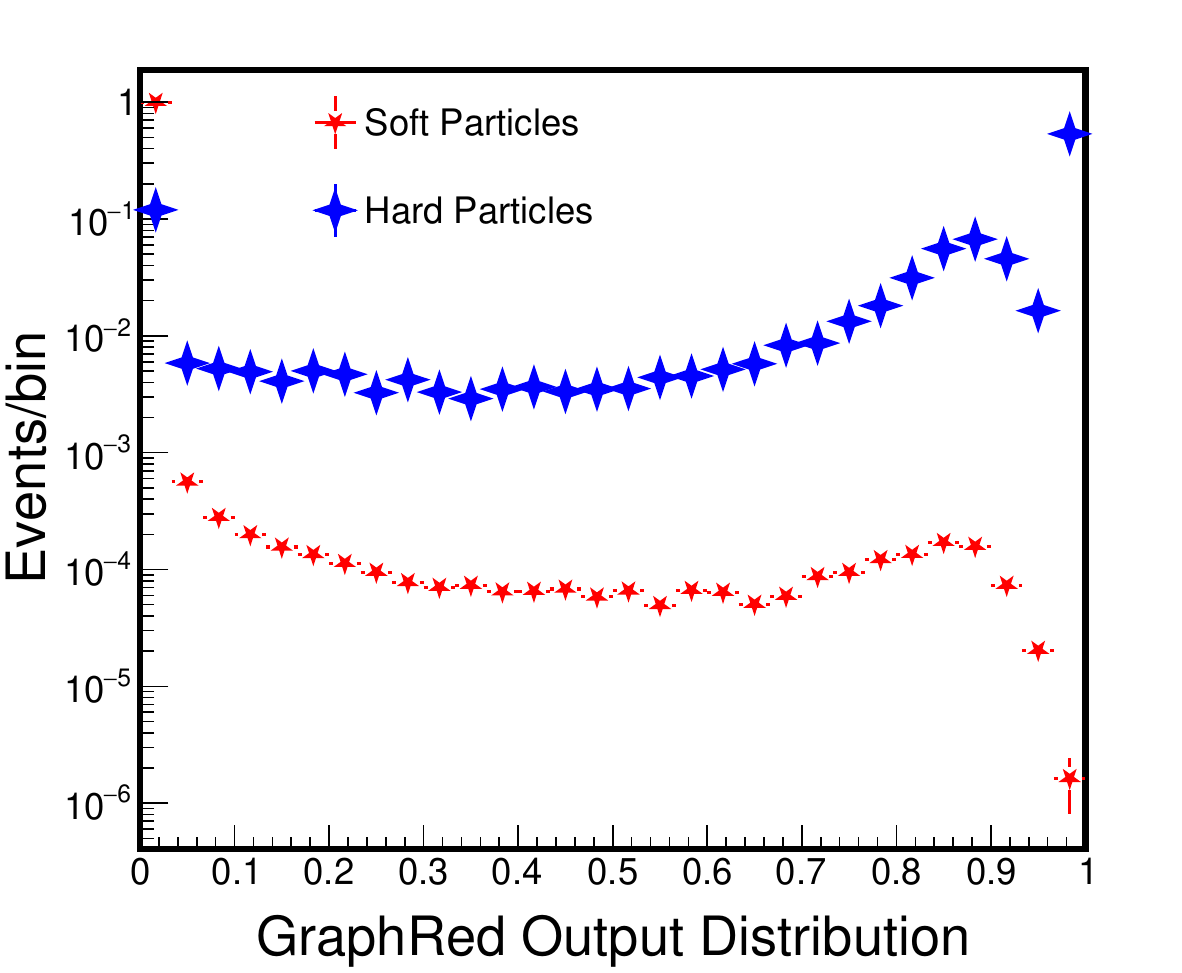}}\\
\end{tabular}
\end{tabularx}
\caption{(color online) GraphRed output probability distribution for hard and soft particles for full heavy-ion event by HYDJET}\label{fig:acc2}
\end{figure}
\newpage
\subsection{Embedded event by PYTHIA $\&$ HYDJET}

The training and validation accuracy w.r.t epochs are shown here for embedded events by PYTHIA $\&$ HYDJET. We infer that an accuracy of $>99\%$ is achieved, describing a good performance of our GraphRed model over the embedded dataset as well.

\begin{figure}[!hbt]
\def\tabularxcolumn#1{m{#1}}
\begin{tabularx}{\linewidth}{@{}cXX@{}}
\begin{tabular}{cc}
\subfloat[Training Accuracy]{\hspace{-1.0em}\includegraphics[scale=0.41]{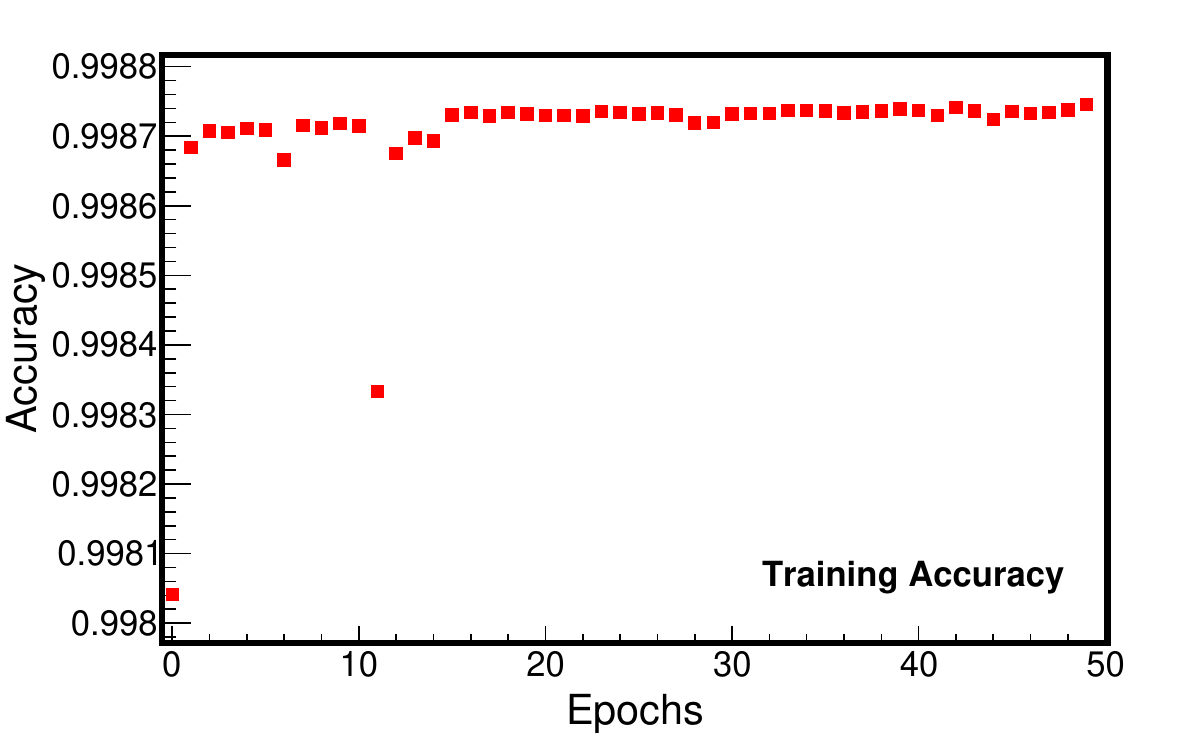}} 
   & \subfloat[Validation Accuracy]{\hspace{-2.0em}\includegraphics[scale=0.41]{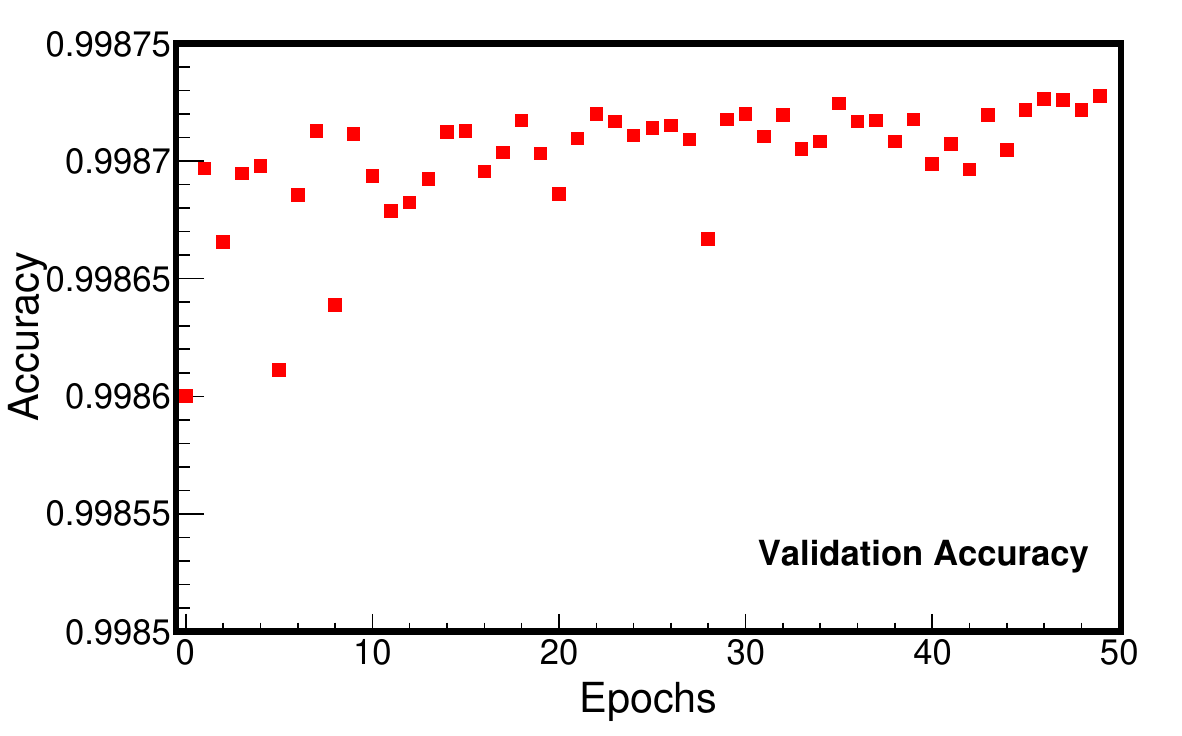}}\\
\end{tabular}
\end{tabularx}
\caption{(color online) Accuracy vs Epochs for Graph Neural Network for embedded events by PYTHIA $\&$ HYDJET}\label{fig:acc_pythia}
\end{figure}
Moreover, the output probability distribution for GraphRed model is shown below for hard and soft particles. We observe that GraphRed is able to identify soft particles depicting a peak around 0 and low miss-classification after 0.7 value. Also, for hard particles there is a peak around 1 describing the correct identification alongside with miss-classification of some particles.
\begin{figure}[!hbt]
    \centering
    \includegraphics[scale=0.45]{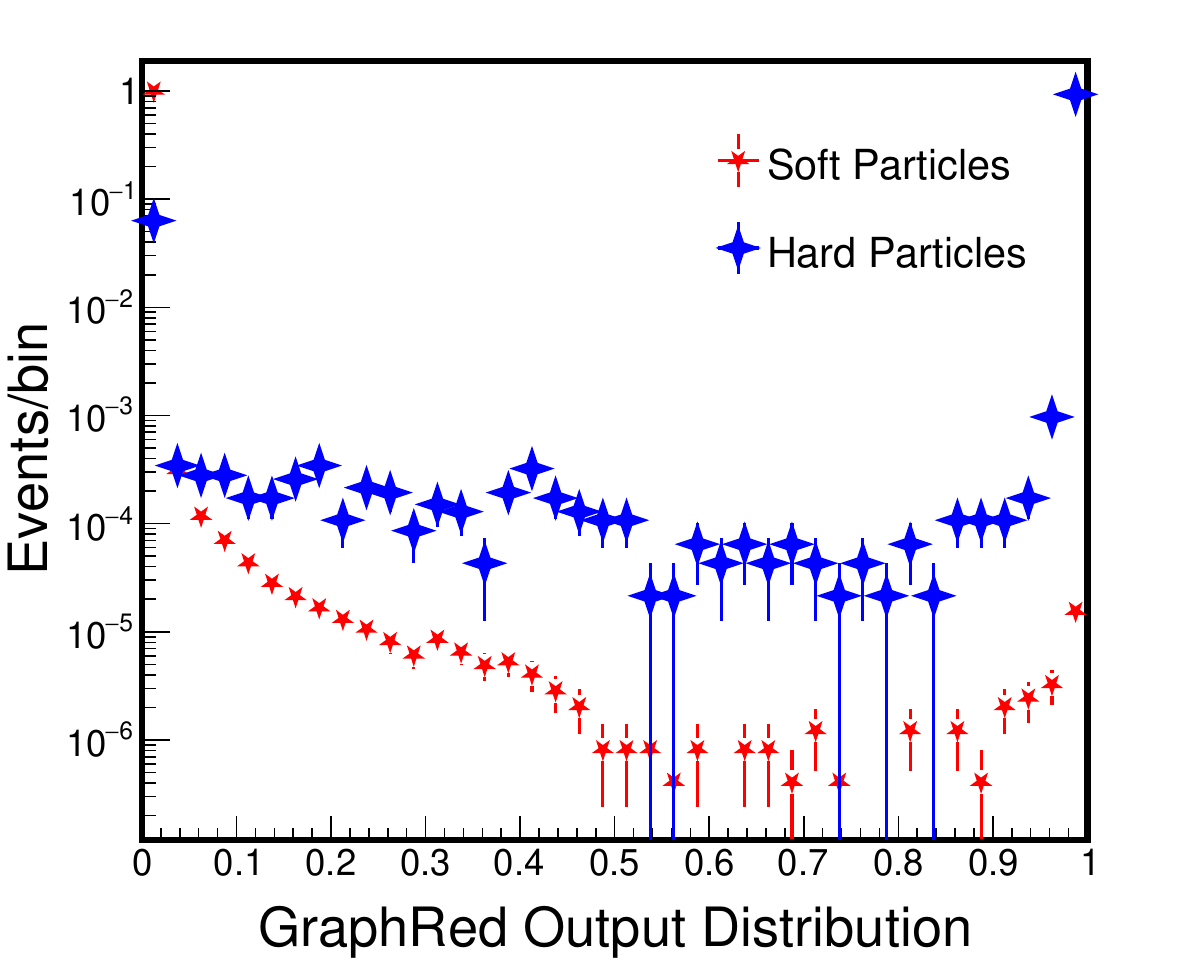}
    \caption{(color online) GraphRed output probability distribution for hard and soft particles for embedded events by PYTHIA $\&$ HYDJET }
    \label{fig:my_label}
\end{figure}


\end{document}